\documentstyle[12pt]{article}
\makeatletter
\@addtoreset{equation}{section}
\makeatother

\newtheorem{theorem}{Theorem} 
\newcommand{\N}{{\mathchoice{\hbox{$\sf\textstyle I\hspace{-.15em}N$}}     
{\hbox{$\sf\textstyle I\hspace{-.15em}N$}}
{\hbox{$\sf\scriptstyle I\hspace{-.10em}N$}}
{\hbox{$\sf\scriptscriptstyle I\hspace{-.11em}N$}}}}
\newcommand{\Z}{{\mathchoice{\hbox{$\sf\textstyle Z\kern-0.4em Z$}}
{\hbox{$\sf\textstyle Z\kern-0.4em Z$}}
{\hbox{$\sf\scriptstyle Z\kern-0.3em Z$}}
{\hbox{$\sf\scriptscriptstyle Z\kern-0.3em Z$}}}}
\newcommand{\R}{{\mathchoice{\hbox{$\sf\textstyle I\hspace{-.15em}R$}}     
{\hbox{$\sf\textstyle I\hspace{-.15em}R$}}
{\hbox{$\sf\scriptstyle I\hspace{-.10em}R$}}
{\hbox{$\sf\scriptscriptstyle I\hspace{-.11em}R$}}}}
\newcommand{\C}{{\mathchoice{\setbox0=\hbox{$\displaystyle\sf C$}\hbox{\hbox
to0pt{\kern0.4\wd0\vrule height0.9\ht0\hss}\box0}}
{\setbox0=\hbox{$\textstyle\sf C$}\hbox{\hbox
to0pt{\kern0.4\wd0\vrule height0.9\ht0\hss}\box0}}
{\setbox0=\hbox{$\scriptstyle\sf C$}\hbox{\hbox
to0pt{\kern0.4\wd0\vrule height0.9\ht0\hss}\box0}}
{\setbox0=\hbox{$\scriptscriptstyle\sf C$}\hbox{\hbox
to0pt{\kern0.4\wd0\vrule height0.9\ht0\hss}\box0}}}}

\renewcommand{\Re}{\mbox{\rm Re }}

\newcommand{\bra}{\langle\,}
\newcommand{\ket}{\,\rangle}

\newcommand{\Tr}{\mbox{\rm Tr}}

\renewcommand{\today}{\number\day \space\ifcase\month\or
  January\or February\or March\or April\or May\or June\or
  July\or August\or September\or October\or November\or December\fi
  \space\number\year}

\newcommand{\cG}{{\cal G}}
\newcommand{\cH}{{\cal H}}
\newcommand{\cU}{{\cal U}}
\newcommand{\ga}{\alpha}
\newcommand{\gb}{\beta}
\newcommand{\gc}{\gamma}
\newcommand{\gC}{\Gamma}
\newcommand{\gd}{\delta}
\newcommand{\gD}{\Delta}
\newcommand{\gep}{\varepsilon}

\newcommand{\gL}{\Lambda}
\newcommand{\gn}{\nabla}
\newcommand{\go}{\omega}
\newcommand{\gO}{\Omega}
\newcommand{\gp}{\varphi}
\newcommand{\gP}{\Phi}
\newcommand{\gs}{\sigma}

\newcommand{\gt}{\theta}
\newcommand{\gT}{\Theta}
\newcommand{\gU}{\Upsilon}
\newcommand{\be}{\begin{eqnarray}}
\newcommand{\ee}{\end{eqnarray}}
\newcommand{\bee}{\begin{eqnarray*}}
\newcommand{\bx}{\bar{x}}

\newcommand{\eea}{\end{array} $$}
\newcommand{\eeaa}{\end{array} \ee}
\newcommand{\eee}{\end{eqnarray*}}

\newcommand{\hh}{\mbox{${{\cal H}}$}}
\newcommand{\hin}{\mbox{${{\cal H}_{in}}$}}
\newcommand{\hout}{\mbox{${{\cal H}_{out}}$}}

\newcommand{\I}{{\rm i}}
\newcommand{\Int}{\displaystyle \int}
\newcommand{\intoi}{\Int_0^\infty}
\newcommand{\intii}{\Int_{-\infty}^{+\infty}}

\newcommand{\intrd}{\Int_{\R^2}}
\newcommand{\Lim}{\displaystyle \lim}
\newcommand{\LL}{ {\textstyle L^2(\frac{dp}{2p},\R_+) } }

\newcommand{\LLw}{ {\textstyle L^2(\frac{d\go}{2\go},\R_+) } }

\newcommand{\Prod}{\displaystyle \prod}

\newcommand{\saut}{\hspace{10mm}}

\newcommand{\sch}{{\cal S}(\R)}
\newcommand{\schd}{{\cal S}(\R^2)}
\newcommand{\sgn}{{\rm sgn}}
\newcommand{\so}{{\cal S}_0(\R)}
\newcommand{\sod}{{\cal S}_0(\R^2)}

\newcommand{\ssp}{{\cal S}(\R_+)}
\newcommand{\SSum}{\displaystyle \sum}

\newcommand{\Trout}{{\Tr \! \! \! \! \!  _{_{_{out}}}} }

\newcommand{\vac}{\gO_{o}}

\newcommand{\abs}[1]{\mid \! #1 \! \mid}

\newcommand{\bea}[1]{$$ \begin{array}{#1}}
\newcommand{\beaa}[1]{\be \begin{array}{#1}}
\newcommand{\ch}[1]{{#1
\hspace{-1.65mm}^{\rule[-2.8mm]{0mm}{4.5mm}{\wedge}} }}

\newcommand{\cha}[1]{{#1
\hspace{-2.1mm}^{\rule[-2.8mm]{0mm}{4.5mm}{\wedge} } }}

\newcommand{\ignore}[1]{}
\newcommand{\ind}[2]{{#1 \! \! _{#2}}}
\newcommand{\indice}[2]{{#1 \! _{#2}}}

\newcommand{\LLu}[1]{{ L^1(d #1,\R) }}

\newcommand{\no}[1]{{ \parallel \! #1 \!\parallel }} 
 
\newcommand{\mt}[1]{{\bra #1 \ket \! _{_\gb}^{Th}}}
\newcommand{\mtbout}[2]{{\bra #1 \ket \! _{_{#2, out}}^{Th}}}

\newcommand{\mtout}[1]{{\bra #1 \ket \! _{_{\gb, out}}^{Th}}}
\newcommand{\rond}[1]{ {#1 
\hspace{-1.3mm}^{\rule[-1.9mm]{0mm}{4mm}{^\circ}} } }
\newcommand{\signe}[1]{{\rm sgn} \left(#1 \right)}
\newcommand{\Signe}[1]{{\rm sgn} \, #1 }
\newcommand{\SS}[2]{ {\rm S}_{#2}{\left[\, #1 \,\right]}}
\newcommand{\sur}[2]{\frac{\textstyle #1}{\textstyle #2}}
\newcommand{\TF}[1]{\widetilde{#1}}
\newcommand{\tf}[1]{{{#1 \! \!}^{^\sim}}}
\newcommand{\tfa}[1]{{{#1 \! \! \!}^{^\sim}}}
\newcommand{\tfch}[1]{{\tfa{\ch{#1}}}}

\newcommand{\tfcha}[1]{{\tfa{\cha{#1}}}}

\newcommand{\wh}[1]{{\widehat{#1}}}
\newcommand{\sub}[1]{\subsection{}{\vspace{-7.2mm}
{\hspace*{3em} \large\bf #1} 
\\ \\ \hspace{-2mm}}}
\begin{document}
\title{A semi-classical relativistic black hole\footnote{
Work done towards a Ph.D.~at Lausanne University. \\
\mbox{\hspace{4.5mm}}$^\dagger$ Leaving for Blackett Laboratory,
Theoretical Physics Group, Imperial College, 
London SW7 2BZ, UK, October 1996.}}
\author{\mbox{}\\ F. Vendrell$^\dagger$ \\ \\
Institut de physique th\'eorique \\
Universit\'e de Lausanne \\
CH-1015 Lausanne, Switzerland\\
E-mail: fvendrel@ipt.unil.ch}
\date{30 July 1996}
\maketitle
\abstract{
A new two-dimensional black hole model, based on the ``$R=T$" 
relativistic theory, is introduced, and the quantum
massless scalar field is studied in its classical 
gravitational field.
In particular infrared questions are discussed.
The two-point function, energy-momentum tensor, current,
Bogoliubov transformations and the mean number of created particles
for a given test function are computed.
I show that this black hole emits massless scalar particles spontaneously.
Comparison with the corresponding field theory in a thermal bath shows 
that the spontaneous emission is {\it everywhere thermal}, i.e. not only 
near the horizon.}
\newpage
\section{Introduction}
S.W.~Hawking discovered that, due to quantum mechanical 
effects, black holes spontaneously create and emit  
particles in 1+3 dimensions.
He showed furthermore that the mean number of spontaneously created 
particles is thermal near the event-horizon \cite{Haw}.
The two-point function and the energy-momentum tensor of quantum 
matter were also computed in the gravitational field of
black holes by other authors and their thermal properties studied 
\mbox{\cite{F2P,TEI}}.
From these results it has been concluded that, near the event-horizon,
the radiation of a 1+3 dimensional black hole is indeed thermal,
with temperature inversely proportional to the mass.

Recently there has been renewed interest in the study of 1+1
dimensional black hole models \cite{CGHS,MST}, for which
the technical difficulties encountered are of less importance than 
in the 1+3 dimensional case.
In the present paper I investigate the semi-classical properties of 
a new 1+1 dimensional black hole model,
based on the ``$R=T$" theory.
This theory was introduced by R.~B.~Mann \cite{Mann}.
The scalar curvature which defines this model vanishes everywhere, 
except on a light-like straight line where it is infinite and from 
which the horizon originates.
I show that this infinite and localized curvature induces an emission 
of massless scalar particles which is thermal everywhere, i.e.~not 
only near the horizon, and that the temperature of the radiation is 
proportional to the relative amplitude of the curvature.

In section 2 the ``$R=T$" theory is reviewed and the new black hole 
model is introduced.
In section 3 the quantization of the massless scalar field 
theory is reviewed in 1+1 dimensional Minkowski space-time.
The quantization is extended to curved space-times in section 4,
where it is also shown that the two-dimensional massless scalar 
field theory may be reduced to two independent one-dimensional scalar
field theories under some specified conditions.
Section 5 is devoted to the formal study of one-dimensional
field theories obtained in this way.
Relevant observables for the massless scalar field are introduced
in section 6.
Section 7 is devoted to the study of one-dimensional massless 
scalar field theories in a thermal bath.
The results obtained are finally applied to the new 
black hole model in section 8.
\section{The relativistic black hole model}
The classical Einstein equations for the gravitational field are 
given by
\be
R_{\mu\nu}-\sur{1}{2}\,g_{\mu\nu} R &=& 8\pi \cG\, T_{\mu\nu},
\label{einstein}
\ee
where $\cG$ is the universal gravity constant and $c=1$.
They imply the covariant conservation of the classical 
energy-momentum tensor $T_{\mu\nu}$:
\be
\gn^\mu \,T_{\mu\nu} &=& 0.
\label{TEIconservation}
\ee
The l.h.s.~of eq.~(\ref{einstein}) vanishes for all 1+1 dimensional 
metrics, so that curvature is arbitrary and matter is excluded 
from 1+1 dimensional space-times \cite{Collas}.
In consequence the Einstein equations have no physical contents in
two dimensions.

In spite of this fact, R.B.~Mann \cite{Mann} has extracted a 
non-trivial theory of gravity from the Einstein equations by 
considering the limit $D\rightarrow 2^+$, where the 
space-time dimension $D$ is allowed to take continuum values. 
The trace of eq.~(\ref{einstein}) is given by
\be
\left(1-\sur{D}{2}\right)\,R(x) &=& 8\pi\cG\,T(x),
\label{gravitybis}
\ee
using $g^{\mu\nu}\,g_{\mu\nu}=D$.
Assuming that the constant $\cG$ depends on the space-time 
dimension $D$ and that the limit
\be
\lim_{D\rightarrow 2^+} \sur{\cG}{1-D/2} &=& G
\ee
exists, then equation (\ref{gravitybis}) implies
\be
R(x) &=& 8\pi G \,T(x),
\label{gravity}
\ee
where $T(x) = T^\mu_{\ \mu}(x)$ is the trace of the energy-momentum 
tensor.
Equation (\ref{gravity}) does not imply the covariant conservation 
of $T_{\mu\nu}(x)$, so eq.~(\ref{TEIconservation}) has to be imposed
by hand.

For the trace $T(x)$ Mann et al.~\cite{MST} have considered the form
\be
T(x) &=& \sur{M}{8\pi G}\,\gd(x^1-x_o^1),
\label{pretrace}
\ee
and have shown that eqs (\ref{gravity}) and (\ref{pretrace}) admit
eternal black holes with a pair of horizons as solutions.

I assume now that $T(x)$ is given by
\be
T(x) &=& \sur{M}{8\pi G}\ \gd(x^+-x_o^+),
\label{pulse}
\ee
where $x^\pm=\left(x^0\pm x^1\right)/\sqrt{2}$ and the constant $M$ is 
strictly positive.
Equation (\ref{pulse}) is consistent with 
eq.~(\ref{TEIconservation}) and describes a pulse of classical 
matter traveling with the velocity of light towards the left at 
$x^+=x^+_o$.
From eqs (\ref{gravity}) and (\ref{pulse}) the scalar curvature 
is given by
\be
R(x) &=& 4M\ \gd(x^+-x_o^+).
\label{modeldef}
\ee
This equation defines a black hole model, as shown below, and
is solved in the conformal gauge
\be
ds^2 &=& C(x)\ dx^+dx^-.
\ee 
Equation (\ref{modeldef}) implies that the conformal factor $C(x)$ 
satisfies the non-linear equation
\be
\partial_+\partial_- \log\vert C(x)\vert 
&=& M\,C(x)\,\gd(x^+-x_o^+).
\ee
This may be rewritten as:
\be
\partial_- \log\vert C(x)\vert &=&
\left\{
\begin{array}{ccl}
C_o, &\hspace{2mm} & \mbox{if \ $x^+< x_o^+$,} \\ [3mm]
M\, C(x^+_o, x^-) + C_o, &&  \mbox{if \ $x^+ > x_o^+$,}
\end{array} \right.
\label{eqdiff}
\ee
where $C_o$ is a real constant, which shows that the 
metric is modified at $x^+=x^+_o$ by the pulse of matter.
This last equation implies that the conformal factor $C(x)$ depends 
only on $x^-$ in the half-plane $x^+ > x_o^+$, and that this is
discontinuous at $x^+=x^+_o$.
This discontinuity comes from the singularity of the curvature 
(\ref{modeldef}) at 
this same value of $x^+$ and it may be removed by replacing the 
delta function (\ref{pulse}) by a sharp continuous pulse centered 
in a neighborhood of $x^+=x^+_o$. 
It is easy to check that a solution of eq.~(\ref{eqdiff}) 
for $C_o=0$ is given by
\be
ds^2 &=& \left\{ \begin{array}{ccl}
dx^+\,dx^-, &\hspace{2mm} & \mbox{if \ $x^+< x_o^+$,} \\ [3mm]
\sur{dx^+\,dx^-}{M\,(\gD-x^-)}, &&  \mbox{if \ $x^+ > x_o^+$,}
\end{array}\right.
\label{metricx}
\ee
where $\gD$ is an arbitrary constant reflecting the invariance
of curvature (\ref{modeldef}) under translations of $x^-$.
Note that to obtain this solution the continuity of $C(x)^{-1}$ has 
been required at $x^-=\gD$ and $x^+ > x_o^+$, where the metric is 
singular.

In a given set of conformal coordinates the {\it horizon} will be 
defined as the curve where the metric reverses its sign.
It thus divides space-time into a {\it time-like} and a 
{\it space-like region}, where the conformal factor is positive
and negative respectively.
The value of the metric may be null or singular on the horizon.
In our case it is singular.
The horizon associated with the metric (\ref{metricx}) is made up of:
%
%
\begin{enumerate}
\item[-] a half-straight line defined by $x^+\geq x^+_o$ and 
$x^-=\gD$ which originates from the singularity of the curvature;
\item[-] a half-straight line defined by 
$x^+=x^+_o$ and $x^-\geq\gD$ superimposed on the singularity of the 
curvature.
\end{enumerate}
The space-like region is identified as the interior of a black
hole, since the events located in it are not in the past
of any observer situated in the flat part of the time-like region 
for all times.
This black hole will be called a {\it relativistic black hole},
because it is based on the relativistic equation (\ref{gravity}).

Since the coordinates $(x^+,x^-)\in\R^2$ are Minkowskian in the 
``past" half-plane $M_P$ defined by (see eq.~(\ref{metricx}))
\be
M_P &=& \{\,x\in\R^2\,\mid\,x^+<x^+_o\,\}, 
\ee
they will be called {\it incoming coordinates}.
Another set of conformal coordinates \mbox{$(y^+,y^-)\in\R^2$} is 
defined by the transformation 
\be
\left\{
\begin{array}{rcl}
x^+(y^+) &=& y^+, \\ [2mm]
x^-(y^-) &=&\gD- e^{-My^-}, 
\end{array}
\right.
\label{transfo}
\ee
which satisfies: 
\be
\lim_{y^-\rightarrow + \infty} x^-(y^-) &=& \gD, \\ [1mm]
\lim_{y^-\rightarrow - \infty} x^-(y^-) &=& -\infty.
\ee
The horizon is located at $y^-=+\infty$ in the new coordinates.
These coordinates cover only the lower part of space-time $R$
defined by
\be
R &=& \{\ x\in\R^2\,\mid\,x^-<\gD\ \},
\ee
where the metric~(\ref{metricx}) is given by 
\be
ds^2 &=& \left\{ \begin{array}{cl}
M\ e^{-My^-}\,dy^+\,dy^-, & \mbox{if \ $y^+< y_o^+$,} \\ [3mm]
dy^+\,dy^-, &  \mbox{if \ $y^+> y_o^+$,} \
\end{array}\right.
\label{metricy}
\ee
where $y^+_o=x^+_o$.
Since the coordinates $(y^+,y^-)$ are Minkowskian in the ``future"
half-plane $M_F$ defined by
\be
M_F &=& \{\,y\in\R^2\,\mid\,y^+>y^+_o\,\}, 
\ee
they will be called {\it outgoing coordinates}.
 
The transformation (\ref{transfo}), which relates incoming 
and outgoing coordinates, is intimately related to the space-time
structure.
It will play an important role in the analysis of the black hole 
semi-classical properties.
Note that the right transformation $x^-(y^-)$ may be extended
analytically in the whole complex plane and that it exhibits 
an imaginary period given by $\frac{2\pi}{M}$ for all the values of 
its argument:
\be
x^-(y^-) &=& x^-\left(y^-+\I\,\sur{2\pi}{M}\,n\right), 
\saut \forall\,n\,\in\Z,\,\forall\,y^-\in\R.
\ee
This period will turn out to be the inverse temperature $\gb$ of 
the black hole radiation.
\section{Quantization of the massless scalar field}
Before considering the quantum physics of the massless scalar field
in 2D curved space-times, its quantization in 2D Minkowski 
space-time should be reviewed.
This cannot be carried out by imposing all the Wightman axioms 
\cite{SW} in a standard way.
In particular the positivity of the Wightman function cannot 
be satisfied for all Schwartz test functions because of its
bad infrared behavior.
Consequently either the massless scalar field should be quantized 
in an indefinite metric following G.~Morchio et al.~\cite{MPS}, or 
the space of test functions should be restricted in order to 
satisfy the positivity condition, as proposed by
S.~Fulling and S.~Ruijsenaars \cite{FR}.
For simplicity I will adopt the second point of view.

In the 2D Minkowski space-time the {\it Wightman distribution} of 
the massless scalar field is defined on the Schwartz space $\schd$ 
by \cite{Wi}
\be
W_o\left[\,h_1\times h_2^*\,\right] &=& 
\intrd d^2k \,\TF{W}_o(k)\ \TF{h}_1(k)\,\TF{h}_2(k)^*, 
\label{Wightman}
\ee
where 
\be
\TF{W}_o(k) &=& \sur{1}{2}\ \left\{\
\gd(k_-)\ \sur{d}{dk_+}\left[\,\gt(k_+)\,\log k_+\,\right]
+ \gd(k_+)\ \sur{d}{dk_-}\left[\,\gt(k_-)\,\log k_-\,\right]
\ \right\}.
\ee
Performing a 2D Fourier transform\footnote{
The 2D Fourier transform is defined by
$\TF{h}(k)=\sur{1}{2\pi}\,
\int_{\R^2} d^2x\,h(x)\,e^{-\I k\cdot x}.$},
the Wightman distribution may also be expressed in the form
\be
W_o\left[\,h_1\times h_2^*\,\right]
&=& \intrd d^2x\intrd d^2\bx\,h_1(x)\,W_o(x,\bx)\,h_2(\bx)^*,
\label{Wightmanx}
\ee
where $W_o(x,\bx)=W_o(x-\bx)$ is the {\it Wightman function} and is
given by
\be
W_o(x) &=& -\sur{1}{8\pi^2}\,\log \left(\,-x^2+\I x^00^+ \,\right) 
- \sur{\gc}{4\pi^2},
\label{W(x)}
\ee
where $\gc$ is the Euler constant.
The Wightman function (\ref{W(x)}) satisfies 
{\it i)} the covariance property, $W_o(\gL x)= W_o(x)$ for any 
Lorentz 
transformation $\gL$; 
{\it ii)} the spectral condition, $\TF{W}_o(k)=0$ if $k^2<0$;
{\it iii)} the locality property, $W_o(x)=W_o(-x)$ if $x^2<0$.
However the positivity condition, $W_o[\,h\times h^*\,]\geq 0$
$\forall\,h\in\schd$, is not generally satisfied (consider 
$\TF{h}(k) = e^{-\ga k^2}$). 
In consequence a standard quantum relativistic interpretation of 
the theory is not possible.

To elude this difficulty, the function space is 
restricted to all Schwartz functions vanishing for null momentum.
The test function space $\sod$ is defined by
\be
\sod &=& \{\,\TF{h}\in\schd \ \mid\ \TF{h}(0)=0\,\},
\ee
and the Wightman distribution (\ref{Wightman}) restricted to 
this space function is given by
\be
W_o[\,h_1\times h_2^*\,] &=&
\intii \sur{dk_1}{2\left\vert k_1\right\vert}\, 
\left[\ \TF{h}_1(k)\ \TF{h}_2(k)^*\,
\right]_{k_o\,=\,\vert k_1\vert},
\label{WightmanR}
\ee
where $\TF{h}_1,\TF{h}_2\in\sod$.
This clearly satisfies the positivity condition and thus
defines a scalar product on $\sod$.
A restricted Hilbert space $\hh$ may now be constructed
from the Wightman distribution (\ref{WightmanR}),
which is related to the two-point function of the scalar field 
$\phi$ by
\be
(\vac, \phi[h_1]\,\phi[h_2]^\dagger\,\vac) &=& 
W_o[\,h_1\times h_2^*\,], 
\label{Wightmanx0}
\ee
where $\vac$ is the vacuum of $\hh$. 

In 2D Minkowski space-time the scalar field $\phi(x)$ satisfies 
the massless Klein-Gordon equation:
\be
\sur{\partial^2\phi}{\partial x^+\,\partial x^-} &=& 0.
\label{KGEx}
\ee
Its general solution will be written in the form
\be
\phi(x) &=& \sur{1}{\sqrt{2\pi}}\,
\left[\ \phi_+(x^+) \ + \ \phi_-(x^-)\ \right],
\label{fieldx}
\ee
where $\phi_+(x^+)$ and $\phi_-(x^-)$ are the left and
right moving fields.
These will be called {\it 1D fields}, in opposition to $\phi(x)$
which is a {\it 2D field}.

The quantum scalar field $\phi$ is defined as a distribution by
\be
\phi[h] &=& \intrd d^2x\,\phi(x)\,h(x),
\label{fieldD}
\ee
where $h$ is any {\it 2D test function} belonging to $\sod$.
The {\it 1D test functions} $h_\pm$ are constructed from the
test function $h$ by integrating on $x^\mp$:
\be
h_\pm(x^\pm) &=& \sur{1}{\sqrt{2\pi}}\,\intii dx^\mp \,\ h(x).
\label{testD1}
\ee
The Fourier transforms of $h$ and $h_\pm$ are related by\footnote{
The 1D Fourier transform is defined by
$ \tfa{h}_\pm(k_\mp) = \frac{1}{\sqrt{2\pi}} 
\int_{\R} dx^\pm\,h_\pm(x^\pm)\ e^{-\I k_\mp x^\pm}. $}
\be
\tfa{h}_\pm(k_\mp) &=& \left.\TF{h}(k) \right\vert_{k_\pm=0},
\label{testFD12}
\ee
and this shows that the functions $\tfa{h}_\pm$
belong to the {\it 1D test function space} $\so$ defined by
\be
\so &=& \{\,\tfa{h}_\pm\in\sch\,\mid\, \tfa{h}_\pm(0)=0\,\},
\ee
if $\TF{h}\in\sod$.
The {\it 1D scalar field distributions} are defined by 
\be
\phi_\pm[h_\pm] &=& \intii dx^\pm\,\phi_\pm(x^\pm)\,h_\pm(x^\pm),
\label{1DfieldDx}
\ee
where $\tfa{h}_\pm\in\so$.
From the previous definitions we deduce that the 2D field 
distribution (\ref{fieldD}) is equal to the sum of the 1D field
distributions (\ref{1DfieldDx}):
\be
\phi[h] &=& \phi_+[h_+] \ + \ \phi_-[h_-].
\label{fieldD12D}
\ee

The {\it 1D Wightman distributions} will be defined on 
$\so\times\so$ by
\be
W_o^\pm[\,h_{1\pm}\times h_{2\pm}^*\,] &=&
\intii dx^\pm \intii d\bar{x}^\pm\
h_{1\pm}(x^\pm)\ W_o^\pm(x^\pm-\bar{x}^\pm)\ 
h_{2\pm}(\bar{x}^\pm)^*,
\label{distributionuni}
\ee
where the {\it 1D Wightman functions}
$W_o^\pm(x^\pm,\bar{x}^\pm)=W_o^\pm(x^\pm-\bar{x}^\pm)$ are 
given, up to a constant, by
\be
W_o^\pm(x^\pm-\bar{x}^\pm)
&=& -\sur{1}{4\pi}\,\log \left(\,\bx^\pm-x^\pm+\I 0^+ \,\right).
\label{Wightmanxunibis}
\ee
From these definitions and eq.~(\ref{W(x)}) we deduce that the 
2D Wightman distribution (\ref{Wightmanx}) is equal to the sum of 
the 1D Wightman distributions (\ref{distributionuni})
\be
W_o[\,h_1\times h_2^*\,] &=& 
W_o^+[\,h_{1+}\times h_{2+}^*\,]
\ + \ W_o^-[\,h_{1-}\times h_{2-}^*\,],
\label{WightmanD12D}
\ee
which are also given by
\be
W_o^\pm[\,h_{1\pm}\times h_{2\pm}^*\,] 
&=& \intoi \sur{dk_\mp}{2k_\mp} \
\tfa{h}_{1\pm}(k_\mp)\ \tfa{h}_{2\pm}(k_\mp)^*,
\label{WightmanxD1D}
\ee
where $\tfa{h}_1,\tfa{h}_2\in\so$.
These are related to the two-point functions by the equations
\be
(\vac, \phi(x)\,\phi(\bx)^\dagger\,\vac) &=& W_o(x-\bx), 
\label{Wightmanx1} \\ [3mm]
(\vac, \phi_\pm(x^\pm)\,\phi_\pm(\bx^\pm)^\dagger\,\vac) &=& 
W_o^\pm(x^\pm-\bx^\pm), 
\label{Wightmanx2} \\ [3mm]
(\vac, \phi_\pm(x^\pm)\,\phi_\mp(\bx^\mp)^\dagger\,\vac) &=& 0,
\label{Wightmanx3}
\ee
from which the fields commutators are computed\footnote{
The equality $2\,\gt(x^2)\ \sgn\, x^0= \sgn\, x^+ + \sgn\, x^-$
is used to obtain eq.~(\ref{commutator2}) from 
eq.~(\ref{commutator1}).}:
\be
\left[\,\phi(x),\,\phi(\bx)^\dagger\,\right] &=& 
\sur{\I}{4\pi}\,\gt[\,(\bx-x)^2\,]\ \Signe (\bx^0-x^0),
\label{commutator1} \\ [2mm]
\left[\,\phi_\pm(x^\pm),\,\phi_\pm(\bx^\pm)^\dagger\,\right] &=& 
\sur{\I}{4}\,\Signe (\bx^\pm-x^\pm), 
\label{commutator2} \\ [2mm]
\left[\,\phi_+(x^+),\,\phi_-(\bx^-)^\dagger\,\right] &=& 0.
\label{commutator3}
\ee

Equations (\ref{fieldD12D}), (\ref{WightmanD12D}), 
(\ref{Wightmanx3}) and (\ref{commutator3}) show 
that the 2D massless scalar field may be considered 
as two uncoupled right and left 1D fields.

We close this section by defining the notion of particle in one
and two dimensions.
These definitions will be useful below.
The function $h\in\sod$ is said to be a 
{\it 2D particle test function} if 
\be
\left.\TF{h}(k)\right\vert_{k_o\,=\,\,-\vert k_1\vert} &=& 0,
\saut  \forall\ k_1\in\R. 
\label{particle2D} 
\ee
Similarly, the functions $h_\pm\in\so$ are said to be 
{\it 1D particle test functions} if 
\be
\tfa{h}_\pm(k^\mp) &=& 0, \saut \forall\ k^\mp<0.
\label{particle1D}
\ee
In the 2D Minkowski space-time eq.~(\ref{testFD12}) implies that 
these definitions are equivalent.
\section{The massless scalar field in curved space-times}
\label{sec:curved}
In this section the field distribution in curved space-times is 
introduced and the relationship between field distributions in 
different coordinates is considered.
The 1D field distributions are defined as in the 2D Minkowski 
space-time, whereas the 1D test functions are defined so as to take 
into account the metric.
I show that, under specified conditions, the relationship between the 
2D field distributions breaks down into two relationships between 1D 
field distributions, so that the 2D quantum problem is reduced to two 
independent 1D quantum problems.
The particle and vacuum concepts are discussed for asymptotically 
Minkowskian coordinates at the end of this section.

I assume that the coordinates $x\in\R^2$ cover a whole
2D space-time.
New coordinates $y$ are introduced by the transformation
\be
y &\longrightarrow& x(y), \saut y\in \R^2,
\label{transfox(y)}
\ee
and they will cover in general only a part $R$ of space-time 
contained in the time-like region.
The scalar fields $\phi(x)$ and $\ch{\phi}(y)$ in these coordinates 
will be called the {\it incoming} and {\it outgoing fields} 
respectively.
They are related by
\be
\ch{\phi}(y) &=& \phi(x(y)),
\saut \forall\,y\in\R^2.
\label{scalarfield}
\ee
The field distributions in both coordinates are defined as follows
\cite{HNS}: 
\be
\phi[h] &=& \int_{\R^2}d^2x\ \sqrt{-g(x)}\ \phi(x)\ h(x),
\label{fielddistributionx} \\ [2mm]
\ch{\phi}[f] &=& 
\int_{\R^2}d^2y\ \sqrt{-\ch{g}(y)}\ \ch{\phi}(y)\ f(y),
\label{fielddistributiony}
\ee
where $h,f\in\sod$.
These definitions are a generalization of eq.~(\ref{fieldD}) to
curved space-times.
The determinants $g(x)$ and $\ch{g}(y)$ of the metric are 
related by
\be
\ch{g}(y) &=&  
\left\vert\,\sur{\partial x}{\partial y}\,(y)\,\right\vert^2\,g(x(y)),
\saut \forall\,y\in\R^2,
\label{transfdeterminant}
\ee
where $\left\vert \partial y/\partial x\right\vert$ is
the Jacobian of the transformation (\ref{transfox(y)}).

Field distributions are considered as geometrical objects whose
values do not depend on the coordinates chosen to express them.
The distributions (\ref{fielddistributionx}) and 
(\ref{fielddistributiony}) are thus related by\footnote{
Note that $f\in\sod$ does not necessarily imply $\ch{f}\in\sod$.
If $\ch{f}\not\in\sod$, eq.~(\ref{fieldtransfo}) is only valid
formally.}
\be
\ch{\phi}[f] &=& \phi[\ch{f}], \saut \forall\,f\in\sod.
\label{fieldtransfo}
\ee
In the region $R$, this last equation defines the {\it incoming test 
function} $\ch{f}(x)$ in terms of the {\it outgoing test function} 
$f(y)$, and I will  assume that $\ch{f}(x)$ vanishes outside the 
region $R$.
Equations (\ref{scalarfield}), (\ref{transfdeterminant}) and 
(\ref{fieldtransfo}) imply that these test functions are related 
in $R$ by
\be
f(y) &=& \ch{f}(x(y)), \saut \forall\,y\in\R^2.
\label{testtransfo}
\ee

Assuming now that the coordinates $x$ are conformal and
that the transformation of coordinates $x=x(y)$ is given by
\be
(y^+, y^-) &\longrightarrow& (x^+(y^+), x^-(y^-)),
\saut (y^+, y^-)\in\R^2,
\label{transfocoordonneesbis}
\ee
then the coordinates $y$ are also conformal.
The property (\ref{transfocoordonneesbis}) is satisfied for
the relativistic black hole model (see eq.~(\ref{transfo})).
In 2D curved space-times, the massless Klein-Gordon equation 
for conformal coordinates is formally identical to the one in
2D Minkowski space-time.
Thus the incoming $\phi(x)$ and outgoing $\ch{\phi}(y)$ fields 
satisfy respectively eq.~(\ref{KGEx}) and
\be
\sur{\partial^2\ch{\phi}}{\partial y^+\,\partial y^-} &=& 0,
\label{KGEy}
\ee
whose solutions are given by eq.~(\ref{fieldx}) and
\be
\ch{\phi}(y) &=& \sur{1}{\sqrt{2\pi}}\,
\left[\ \indice{\ch{\phi}}{+}(y^+) \ + \ 
\indice{\ch{\phi}}{-}(y^-)\ \right].
\label{fieldy}
\ee
The relation between the left and right fields is deduced
from eq.~(\ref{scalarfield}) up to a constant:
\be
\ch{\phi}_\pm (y^\pm) &=& \phi_\pm (x^\pm(y^\pm)),
\saut \forall\,y^\pm\in\R.
\label{fieldxy1D}
\ee

In 2D curved space-times the {\it 1D test functions} are defined by
\be
h_\pm(x^\pm) &=& 
\sur{1}{\sqrt{2\pi}}\,\intii dx^\mp \,\sqrt{-g(x)}\ h(x),
\label{1Dtestx}\\ [1mm]
f_\pm(y^\pm) &=& 
\sur{1}{\sqrt{2\pi}} \intii dy^\mp \,\sqrt{-\ch{g}(y)}\ f(y).
\label{1Dtesty}
\ee
These definitions include the determinant of the metric and are a 
generalization of eq.~(\ref{testD1}).
The {\it 1D incoming and outgoing field distributions} are defined as
in Minkowski space-time and are given by 
eq.~(\ref{1DfieldDx}) and
\be
\ch{\phi}_\pm[f_\pm] 
&=& \intii dy^\pm\,\phi_\pm(y^\pm)\,f_\pm(y^\pm).
\label{1DfieldDy}
\ee 
Equation (\ref{fieldD12D}) is still valid in 2D curved space-times in 
the $x$ and $y$ coordinates:
\be
\phi[h] &=& \phi_+[h_+] \ + \ \phi_-[h_-], 
\label{2D1fieldx} \\ [3mm]
\ch{\phi}[f] &=& \indice{\ch{\phi}}{+}[f_+]
\ + \ \indice{\ch{\phi}}{-}[f_-].
\label{2D1fieldy}
\ee
The transformations for the  1D test functions are deduced from 
eqs (\ref{testtransfo}), (\ref{1Dtestx}) and (\ref{1Dtesty}):
\be
f_\pm(y^\pm) &=& 
\sur{\partial x^\pm}{\partial y^\pm}\,(y^\pm)\
\ind{\ch{f}}{\pm} (x^\pm(y^\pm)),
\saut y^\pm\in\R.
\label{1Dtestfunctiontransfo}
\ee
The metric does not appear explicitly in these transformations
although they contain the dynamics of the problem.
They imply that the 2D field transformation (\ref{fieldtransfo}) may 
be broken down into two 1D left and right field transformations:
\beaa{rclcrcl}
\ch{\phi}_+[f_+] &=& \phi_+\,[\ind{\ch{f}}{+}], 
&\saut&
\ch{\phi}_-[f_-] &=& \phi_-\,[\ind{\ch{f}}{-}].
\label{leftrightfieldtransfo}
\eeaa
I must emphasize that the 1D field distributions $\ch{\phi}_\pm$ and 
$\phi_\pm$ are formally identical with their Minkowskian counterparts.
Equations (\ref{leftrightfieldtransfo}) imply that the left and right
modes of the fields are not mixed up by changing coordinates.
They are thus dynamically independent.

Note that the definitions (\ref{particle2D}) and (\ref{particle1D}) 
for 2D and 1D particle test functions are not {\it strictly} 
equivalent in curved space-times in any coordinates 
(see eq.~(\ref{1Dtestx}) or (\ref{1Dtesty})).
There may however be {\it approximate} equivalence if the 2D test 
function is ``well localized"\footnote{
Note that a 2D particle test function cannot in general be strictly 
localized, since its Fourier transform does not contain negative
contributions.} 
in a space-time region $M$ where the metric is (asymptotically) 
Minkowskian.
This shows that it is difficult to give a precise meaning
to the notion of particle in curved space-time and in particular to 
make this meaning coincident with that of the Minkowskian field 
theory.

We note furthermore that the notions of particle are different in
the $x$ and $y$ coordinates.
In the 1D language, the particles test functions are defined 
respectively by
\be
\tfa{h}_\pm (k^\pm) &=& 0, \saut \mbox{if $k^\mp<0$}, 
\label{eq:incomingparticle}\\ [2mm]
\ind{\tf{f}}{\pm} (p^\pm) &=& 0, \saut \mbox{if $p^\mp<0$}.
\label{eq:outgoingparticle}
\ee
These conditions are incompatible unless the transformation
$x(y)$ is the identity, i.e.~the scalar curvature vanishes everywhere.
This incompatibility is the key to understanding the creation of 
particles in curved space-times.

We assume from now on that the coordinates $x$ and $y$ are 
(asymptotically) Minkowskian in past and future space-time regions 
$M_P$ and $M_F$ respectively (as is the case in the relativistic
black hole model).
In consequence, they will be called {\it incoming} and {\it outgoing
coordinates} respectively.
If the test functions $h(x)$ and $f(y)$ are well localized in $M_P$ 
and $M_F$, and satisfy respectively eqs (\ref{eq:incomingparticle})
and (\ref{eq:outgoingparticle}), then they will respectively describe 
{\it incoming} and {\it outgoing particles}. 

The {\it incoming and outgoing vacuums}, $\vac$ and $\Psi_o$, will be
defined in the 1D language by
\be
\phi_\pm[h_\pm]\,\vac &=& 0,
\label{1Dvacuumin} \\ [2mm]
\ch{\phi}_\pm[f_\pm]\,\Psi_o &=& 0,
\label{1Dvacuumout}
\ee
where $h_\pm(x^\pm)$ and $f_\pm(y^\pm)$ are arbitrary 1D 
particle test functions (i.e.~they satisfy respectively eqs 
(\ref{eq:incomingparticle}) and (\ref{eq:outgoingparticle})).
Furthermore, if the corresponding 2D test functions $h(x)$ and $f(y)$ 
are also well localized in $M_P$ and $M_F$ respectively, these 
equations imply from eqs (\ref{2D1fieldx}) and (\ref{2D1fieldy})
\be
\phi[h]\,\vac &\approx& 0,
\label{2Dvacuumin} \\ [2mm]
\ch{\phi}[f]\,\Psi_o &\approx& 0,
\label{2Dvacuumout}
\ee
and the functions $h(x)$ and $f(y)$ are 2D particle test functions 
(i.e.~$h(x)$ satisfies eq.~(\ref{particle2D}) in the incoming 
coordinates and $f(y)$ satisfies a similar equation in the outgoing 
coordinates).
We thus conclude that the vacuums $\vac$ and $\Psi_o$ are ordinary 
Minkowskian vacuums.
In particular, the incoming vacuum $\vac$ is formally equivalent to 
the vacuum of the preceding section and consequently 
eqs (\ref{Wightmanx1}) to (\ref{Wightmanx3}) for the two-point 
functions are also valid in curved space-times.
\section{One-dimensional scalar field theory}
In this section the one dimensional scalar field theories are 
studied.
I show that the commutation relations of the fields are invariant
under any change of coordinates.
The Bogoliubov transformations between the incoming and outgoing
field operators are obtained and their implementability
is discussed.
Note that, for the relativistic black hole model, the physics of 
the left moving field $\phi_+$ is trivial, since the transformation 
(\ref{transfo}) between the left coordinates is the identity.
I shall consider from now on only the right moving 
field $\phi_-$ and shall drop the subscript $-$. 

The scalar product $\bra\ ,\ \ket$ of two test functions
is given by (see eq.~(\ref{WightmanxD1D}))
\be
\bra \ind{\tf{f}}{2},\ind{\tf{f}}{1} \ket
&=& \intoi \sur{dp}{2p} \ 
\ind{\tf{f}}{2}(p)^* \,\ind{\tf{f}}{1}(p),
\ee
where $\ind{\tf{f}}{1},\ind{\tf{f}}{2}\in\so$.
The norm $\no{\mbox{\ }}$ is defined by
\be
\no{\tf{f}}^2 &=& \bra \tf{f},\tf{f} \ket.
\ee
We define furthermore the function spaces
\be
\ssp &=& \{\ \tf{f}\in \so \mid \ 
\tf{f}(p)=\gt(p)\,\tf{f}(p) \ \ \forall\,p\in\R\ \}, \\ [2mm]
\LL &=& \{\ \tf{f} \ \mid\  
\tf{f}(p)=\gt(p)\,\tf{f}(p) \ \ \forall\,p\in\R \ \
\mbox{and} \ \ \no{\tf{f}} <\infty \ \}.
\ee
Note that
\be
\overline{\ssp}^{\ \no{\ \ }}&=& \LL.
\ee
The set $\so$ is the {\it particle test function space}
and $\LL$ is the {\it particle wave function space}.

We recall that the incoming and outgoing test functions
are related by (see eq.~(\ref{1Dtestfunctiontransfo}))
\be
f(y) &=& \sur{\partial x}{\partial y}\,(y)\ \ch{f} (x(y)),
\label{fTransformation}
\saut \forall\,y\in\R.
\ee
It is not clear whether the inverse Fourier transform $f(y)$ and
the Fourier transform $\tfch{f}(k)$ exist if $\tf{f}\in\LL$.
For simplicity, I will assume in the following that $f(y)$ exists 
a.e.~and is integrable, so that the existence 
of $\tfch{f}(k)$ is certain.
Note that $f\in \LLu{y}$ implies that $\ch{f}(x)$ is also 
integrable.
This hypothesis is thus formulated in a way which is invariant
under any transformation of coordinates.
It also implies that the Fourier transforms $\tf{f}(p)$ and
$\tfch{f}(k)$ are continuous everywhere and vanish at infinity.
The incoming and outgoing momenta will be denoted by $k$ and $p$ 
respectively.

The Fourier transforms of the incoming and outgoing wave functions
will be related by the operator $U$ defined by
\be
\tfch{f}(k) &=& \intoi dp\ U(k,p)\,\tf{f}(p),
\ee
whose kernel $U(k,p)$ is given by
\be
U(k,p) &=& \sur{1}{2\pi}\,\intii dy\ e^{-\I kx(y)}\,e^{\I py}.
\label{KernelU}
\ee 
For any transformations $x=x(y)$, this satisfies the property
\be
U(0,p) &=& \gd(p), \saut \forall\,p\in\R_+,
\label{Udelta}
\ee
which implies $\tfch{f}(0)=\tf{f}(0)=0$ under our assumptions.

The positive and negative momentum components of the outgoing
and incoming test functions $\tf{f}(p)$ and $\tfch{f}(k)$ are 
defined as
\beaa{rclcrcl}
\ind{\tf{f}}{_P} (p) &=& \gt(p) \, \tf{f}(p), &\saut&
\ind{\tf{f}}{_N} (p) &=& \gt(p) \, \tf{f}(-p), \\ [2mm]
\ind{\tfch{f}}{_P} (p) &=& \gt(k) \, \tfch{f}(k), &\saut&
\ind{\tfch{f}}{_N} (p) &=& \gt(k) \, \tfch{f}(-k).
\eeaa
The operators $A$ and $B$ will be defined respectively as the 
positive and negative incoming momentum contributions of $U$
\beaa{rclcrcl}
(A\tf{f})(k) &=& (U\tf{f})_{_P}(k), &\saut& 
(B\tf{f})(k) &=& (U\tf{f})_{_N}(k),
\label{defAB}
\eeaa
and the bilinear operator $G$ by
\be 
G\, (f_1\times f_2) &=& -\frac{1}{4\pi} \,
\intii dy \intii dy' \, f_1(y)\, 
\log \left[\,\sur{x(y)-x(y')}{y-y'}\,\right]\,f_2(y').
\label{defG}
\ee
The logarithm in the integrand of this double integral is well 
defined since $x(y)$ is always an increasing function.
The scalar product of incoming functions such as (\ref{defAB}) 
may be expressed in terms of the bilinear operator $G$ 
evaluated for the corresponding outgoing functions, as shown in 
the following theorem.
\begin{theorem} \label{th:formulae}
If $\ind{\tf{f}}{1},\ind{\tf{f}}{2} \in \LL$ are two wave functions 
such that their inverse Fourier transforms exist and are 
integrable, then
\be
\bra A\ind{\tf{f}}{2},A\ind{\tf{f}}{1} \ket 
&=& G\ (f_1 \times f_2^*) \ + \
\bra \ind{\tf{f}}{2},\ind{\tf{f}}{1} \ket ,
\label{formulaeA} \\ [1mm]
\bra B\ind{\tf{f}}{2},B\ind{\tf{f}}{1} \ket 
&=& G\ (f_1 \times f_2^*) ,
\label{formulaeB} \\ [1mm]
\bra A^*\ind{\tf{f}}{2}^*,B\ind{\tf{f}}{1} \ket 
&=& G\ (f_1 \times f_2), 
\label{formulaeC} \\ [1mm]
\bra B^*\ind{\tf{f}}{2}^*,A\ind{\tf{f}}{1} \ket 
&=& G\ (f_1 \times f_2),
\ee 
and hence\footnote{
Equations (\ref{AABBE}) and (\ref{ABBA}) were first obtained
by R.~M. Wald \cite{Wal}.}
\be
A^\dagger A &=& B^\dagger B + E, 
\label{AABBE} \\ [1mm]
A^T B &=& B^T A,
\label{ABBA}
\ee
where $E$ is the identity.
\end{theorem}
Equation (\ref{formulaeB}) is proved in appendix 
\ref{ap:formulaeB} and the others results of this theorem 
are proved in a similar way.

We recall that the incoming and outgoing fields are related by
(see eq.~(\ref{fieldxy1D}))
\be
\ch{\phi}(y) &=&\phi(x(y)), \saut \forall\,y\in\R.
\label{fieldxy1Dbis}
\ee
The Wightman function for the incoming fields is given by the 
equation (see eq.~(\ref{WightmanxD1D}))
\be
(\vac, \phi[h_1]\,\phi[h_2]^\dagger\,\vac) &=&
\intoi \sur{dk}{2k} \ \tfa{h}_2(k)^* \,\tfa{h}_1(k),
\label{one2pf}
\ee
from which their commutator is deduced\footnote{
Equation (\ref{commutator}) may also be obtained from the 
commutator (\ref{commutator2}) using 
$\sur{i}{\pi}\,\int_{-\infty}^{+\infty} 
dk\,P\sur{1}{k}\,e^{-\I kx} = \sgn\,x$.}
\be
\left[\ \phi [h_1], \, \phi [h_2]^\dagger\ \right]
&=& \intii \sur{dk}{2k} \ \tfa{h}_2(k)^* \,\tfa{h}_2(k),
\label{commutator}
\ee
if $\tfa{h}_1,\tfa{h}_2\in\so$.
We have a similar result for the outgoing fields.
The equality
\be
\intii\sur{dk}{2k}\,\ind{\tfch{f}}{2}(k)^*\,\ind{\tfch{f}}{1}(k) 
&=&\intii\sur{dp}{2p}\,\ind{\tf{f}}{2}(p)^*\,\ind{\tf{f}}{1}(p),
\label{commutatorequality}
\ee
proved in appendix \ref{ap:commutatorequality},
implies that the field commutator is invariant under any 
transformation of coordinates $x=x(y)$
\be
\left[\ \ch{\phi} [f_1],\, \ch{\phi} [f_2]^\dagger\ \right]
&=& \left[\ \phi [f_1], \, \phi [f_2]^\dagger\ \right],
\label{commutatorinvariance}
\ee
where $\ind{\tf{f}}{1},\ind{\tf{f}}{2} \in \so$.
The Wightman function (\ref{one2pf}) is, however, not invariant
under any non-trivial transformation of coordinates.

In the {\it real} scalar fields,
the incoming and outgoing field operators $a_{in,out}$ and 
$a_{in,out}^\dagger$ are defined by splitting the positive and 
ne\-gative momentum contributions of the incoming and outgoing 
fields:
\be
\phi [h] &=& a_{in}[\indice{\tfa{h}}{_P}] 
\ + \ a_{in}^\dagger[\indice{\tfa{h}}{_N}], 
\label{fieldopin}\\ [2mm]
\ch{\phi} [f] &=& a_{out}[\ind{\tf{f}}{_P}] 
\ + \ a_{out}^\dagger[\ind{\tf{f}}{_N}],
\label{fieldopout}
\ee
and they are
annihilation and creation operators respectively.
By applying the incoming and outgoing creation operators 
respectively on the
vacuums $\vac$ and $\Psi_o$ (see def.~(\ref{1Dvacuumin}) and 
(\ref{1Dvacuumout})), the {\it Hilbert spaces} $\hin$ and $\hout$ 
are constructed.
The incoming and outgoing field operators are related by
\beaa{rcccccl}
a_{out}[\tf{f}] &=& \ch{\phi} [f] &=& \phi [\ch{f}] 
&=& a_{in}[(U\tf{f})_{_P}] \ + \ a_{in}^\dagger[(U\tf{f})_{_N}],
\eeaa
if $\tf{f}\in\so$, and the {\it Bogoliubov transformations} are 
thus given by
\beaa{rcl}
a_{out}[\tf{f}] &=& a_{in}[A\tf{f}] \ + \ a_{in}^\dagger[B\tf{f}], 
\\ [3mm]
a_{out}^\dagger[\tf{f}] &=& 
 a_{in}[B^*\tf{f}] \ + \ a_{in}^\dagger[A^*\tf{f}].
\label{bogoliubov}
\eeaa

Since $\phi [h] = a_{in}[\tfa{h}]$ if $\tfa{h}\in\ssp$,
we deduce from  eq.~(\ref{commutator}) that the field operator 
commutators are
\beaa{rclcl}
\left[ \
a_{in}[\tfa{h}_1], \, a_{in}[\tfa{h}_2]^\dagger \ \right]
&=& \bra \tfa{h}_2, \tfa{h}_1 \ket, &\hspace{27mm}&
\label{commutatorin} 
\eeaa
\mbox{}\vspace{-10mm}\\
\beaa{rclcl}
\left[ \ a_{in}[\tfa{h}_1],\, a_{in}[\tfa{h}_2] \ \right]
&=&  
\left[ \
a_{in}[\tfa{h}_1]^\dagger,\, a_{in}[\tfa{h}_2]^\dagger \ \right]
&=&0,
\label{commutatorinbis}
\eeaa
where $\tfa{h}_1,\,\tfa{h}_2\in\ssp$.
From the invariance of the field commutator 
(\ref{commutatorinvariance}), it is clear that the field operator 
commutators are also invariant:
\be
\left[\ a_{in} [\tfa{h}_1], \, 
a_{in} [\tfa{h}_2]^\dagger\ \right] &=&
\left[\ a_{out}[\tfa{h}_1],\, a_{out}[\tfa{h}_2]^\dagger\ \right],
\label{opcommutatorinvariance}\\ [3mm]
\left[\ a_{in} [\tfa{h}_1], \, 
a_{in} [\tfa{h}_2] \ \right] &=&
\left[\ a_{out}[\tfa{h}_1],\, a_{out}[\tfa{h}_2]\ \right],
\label{opcommutatorinvariancebis}
\ee
where $\tfa{h}_1,\,\tfa{h}_2\in\ssp$.
Note that eqs (\ref{commutatorin}) to 
(\ref{opcommutatorinvariancebis}) also imply the fundamental 
relations (\ref{AABBE}) and (\ref{ABBA}).

The field operator modes $a_{out}(p)$ and $a_{in}(k)$ are defined 
by
\beaa{rclcrcl}
a_{out}[\tf{f}] &=& 
\intoi \sur{dp}{2p} \ a_{out}(p) \ \tf{f}(p), &\hspace{2mm}&
a_{in}[\tfa{h}] &=& 
\intoi \sur{dk}{2k} \ a_{in}(k) \ \tfa{h}(k),
\eeaa
where $\tfa{h},\tf{f}\in\ssp$.
Expansions (\ref{fieldopin}) and (\ref{fieldopout}) are rewritten
as
\be
\phi(x) &=& \sur{1}{\sqrt{2\pi}} \,
\intoi \sur{dk}{2k} \, \left[\, 
a_{in}(k)\,e^{-\I kx} + a_{in}^\dagger(k)\,e^{\I kx}\,\right],
\label{devphiin} \\ [2mm]
\ch{\phi}(y) &=& \sur{1}{\sqrt{2\pi}} \,
\intoi \sur{dp}{2p} \, \left[\, 
a_{out}(p) \, e^{-\I py} + a_{out}^\dagger(p)\,e^{\I py}\,\right].
\label{devphichapout}
\ee
These are representations of the fields $\phi(x)$ and 
$\ch{\phi}(y)$ in the Hilbert spaces $\hin$ and $\hout$ 
respectively.
The representation of the outgoing field $\ch{\phi}(y)$ in the 
incoming Hilbert space $\hin$ is deduced from eqs 
(\ref{fieldxy1Dbis}) and (\ref{devphiin}):
\be
\ch{\phi}(y) = \sur{1}{\sqrt{2\pi}} \,
\intoi \sur{dk}{2k} \, \left[\, 
a_{in}(k) \, e^{-\I kx(y)} + a_{in}^\dagger(k) \, e^{\I kx(y)}
\,\right].
\label{devphichapin}
\ee

The operator $V$ is defined by the kernel
\be
V(k,p) &=& \sur{1}{2\pi}\, \int_I dx\, e^{-\I kx}\,e^{\I py(x)},
\label{kernelV}
\ee 
where $I=\{\,x(y) \ \vert\ y\in\R \,\}$.
The operators $U$ and $V$ satisfy the properties
\be
V(k,p) &=& \sur{p}{k}\ U(k,p), \saut \forall\,k,\,p\in\R,
\label{eq:applicationVA} \\ [1mm]
V^\dagger\,U &=& E,
\label{eq:applicationVB} \\ [2mm]
U\,V^\dagger = E &\Longleftrightarrow & I=\R,
\label{eq:applicationVC}
\ee
where $E$ is the identity operator.
Thus $U$ is non-singular if and only if $I=\R$,
and if $I=\R$, we have $U^{-1}=V^\dagger$.

Using the kernel (\ref{kernelV}), the Bogoliubov transformations 
(\ref{bogoliubov}) may be rewritten in the form
\be
\left( 
\begin{array}{c} a_{out}(p) \\ a_{out}^\dagger(p)\end{array}
\right)
&=& \intoi dk\ \left( \begin{array}{cc} 
         V(k,p) &  -V(-k,p) \\
        -V(k,-p) &  V(-k,-p)
         \end{array} \right) \
\left( \begin{array}{c} 
       a_{in}(k) \\ a_{in}^\dagger(k)
       \end{array} \right).
\label{Bogoliubovbis}
\ee
Equation (\ref{eq:applicationVC}) implies that the Bogoliubov 
transformations (\ref{bogoliubov}) and (\ref{Bogoliubovbis}) are 
invertible if and only if $I=\R$.

Assuming now that $I=\R$, we split the outgoing positive and 
negative momentum contributions of $V^\dagger\tfa{h}$, defining 
the operators $C$ and $D$ by
\beaa{rclcrcl}
(C\tfa{h})(p) &=& (V^\dagger\tfa{h})_{_P}(p), &\saut&
(D\tfa{h})(p) &=& (V^\dagger\tfa{h})_{_N}(p).
\eeaa
The inverse of the Bogoliubov transformation (\ref{bogoliubov}) 
is then given by
\beaa{rcl}
a_{in}[\,\tfa{h}\,] &=& 
a_{out}[\,C\tfa{h}\,] \ + \ a_{out}^\dagger[\,D\tfa{h}\,], 
\\ [3mm]
a_{in}^\dagger[\,\tfa{h}\,] &=& 
a_{out}[\,D^*\tfa{h}\,] \ + \ a_{out}^\dagger[\,C^*\tfa{h}\,],
\label{bogoliubovinvers}
\eeaa
if $\tfa{h}\in\ssp$, or by
\be
\left( 
\begin{array}{c} a_{in}(k) \\ a_{in}^\dagger(k)\end{array}
\right)
&=& \intoi dp\ \left( \begin{array}{cc} 
          U(-k,-p) &  -U(-k,p) \\
          -U(k,-p) &  U(k,p)
         \end{array} \right) \
\left(   \begin{array}{c} 
          a_{out}(p) \\ a_{out}^\dagger(p)
         \end{array} \right).
\ee

The previous results are easily generalized to the {\it complex} 
scalar field, for which the field operators $a_{in,out}$ and
$b_{in,out}^\dagger$ are defined by
\be
\phi [h] &=& a_{in}[\tfa{h}_{_P}] 
\ + \ b_{in}^\dagger[\tfa{h}_{_N}], \\ [2mm]
\ch{\phi} [f] &=& a_{out}[\ind{\tf{f}}{_P}] 
\ + \ b_{out}^\dagger[\ind{\tf{f}}{_N}],
\label{fieldoperatorscomplex}
\ee
if $\tfa{h},\tf{f}\in\so$.
The representation of the complex scalar field $\ch{\phi}(y)$ 
in the Hilbert space $\hin$ is given by
\be
\ch{\phi}(y) &=& \sur{1}{\sqrt{2\pi}} \,
\intoi \sur{dk}{2k} \, \left[\, 
a_{in}(k) \, e^{-\I kx(y)} + b_{in}^\dagger(k) \, e^{\I kx(y)}
\,\right],
\label{fieldincomplex}
\ee
and the relations
\beaa{rcl}
a_{out}[\tf{f}] &=& a_{in}[A\tf{f}] \ + \ b_{in}^\dagger[B\tf{f}], 
\\ [3mm]
b_{out}^\dagger[\tf{f}] &=& 
 a_{in}[B^*\tf{f}] \ + \ b_{in}^\dagger[A^*\tf{f}],
\label{bogoliubovcomplex}
\eeaa
where $\tf{f}\in\ssp$, are the associated Bogoliubov 
transformations.

The outgoing test functions $\ind{\tf{f}}{p_o}$ of mode $p_o$ are 
defined formally as
\be
\ind{\tf{f}}{p_o}(p) &=& 2p\ \gd(p-p_o).
\label{defmodep}
\ee
This definition is correct only if $p_o>0$.
The null mode $\ind{\tf{f}}{p_o\,=\,0}$ is defined as the
limit $n\rightarrow \infty$ of the series \cite{MPS}
\be
\ind{\tf{f}}{_0}^{(n)}(p) &=& 
\sur{\tfa{h}_n(p)}{\bra \tfa{h}, \tfa{h}_n\ket},
\label{eq:series}
\ee
where $\tfa{h}(p)=e^{-p^2}$ and 
$\tfa{h}_n(p)= \tfa{\chi}(np)\ \tfa{h}(p)$,
with the function $\tfa{\chi}$ defined by
\beaa{rcccccccl}
0 &\leq& \tfa{\chi}(p) &\leq& 1, \hspace{3mm} \forall \, p\,\in\R,
&\hspace{3mm} \mbox{and} \hspace{3mm}&
\tfa{\chi}(p) &=& \left\{
\begin{array}{cl}
0, & \mbox{if \ $p\leq 0$,} \\ [2mm]
1, & \mbox{if \ $p\geq 1$.}
\end{array} \right.
\label{chi}
\eeaa
The series (\ref{eq:series}) satisfies
\be
\lim_{n\rightarrow \infty} \bra \ind{\tf{f}}{_0}^{(n)},\tf{f} \ket
&=& \tf{f}(0),
\saut \mbox{if $\tf{f}\in\sch$},
\\ [1mm]
\lim_{n\rightarrow \infty} \no{\ind{\tf{f}}{_0}^{(n)}}&=&0.
\ee

The generalized functions $f_{p_o}$ ($p_o\geq 0$) are not 
normalizable and thus they are not associated to a state in the 
Hilbert space $\hout$.

Let $\{ \ind{\tf{f}}{i} \}_{i=1}^n \subset \ssp$ be a set of 
normalized particle test functions.
The {\it $n$-particle test function} $f^{(n)}$ is defined as
\be
f^{(n)} &=& C \, f_1 \times f_2 \times ... \times f_n,
\label{deffbar}
\ee
where $\times$ is the tensor product and $C$ a constant.
A product of fields is also defined,
\be
\ch{\phi}[f^{(n)}] &=& C \, \ch{\phi}[f_1] \ \ch{\phi}[f_2]
\  ... \ \ch{\phi}[f_n],
\ee
and the state denoted $\Psi_{f^{(n)}}$ is given in terms
of this product by
\be
\Psi_{f^{(n)}} &=& \ch{\phi}[f^{(n)}]^\dagger \, \Psi_o.
\ee
The state $\Psi_{f^{(n)}}$ is normalized by imposing the equation
\beaa{rcccl}
(\Psi_{f^{(n)}}, \Psi_{f^{(n)}}) &=&
(\Psi_o,\ch{\phi}[f^{(n)}]\ \ch{\phi}[f^{(n)}]^\dagger\,\Psi_o)
&=& 1,
\label{fbarnormal}
\eeaa
which fixes the constant $C$.
\section{Observables in the outgoing coordinates}
In this section mean values of observables, built into the 
outgoing coordinates, are computed in the incoming vacuum.
These quantities describe the properties of the outgoing
particles created by the space-time curvature.
The two-point function, energy-momentum tensor, 
current for the complex scalar field and the mean number
of spontaneously created particles for a given outgoing test 
function are considered.
The total mean number of particles is computed and the
implementability of $U$ is also considered.

The {\it outgoing two-point function} $\wh{W}_o(y,y')$ is defined 
as the mean value of outgoing fields in the incoming vacuum:
\be
\wh{W}_o(y,y') 
&=& (\vac,\ch{\phi}(y)\,\ch{\phi}(y')^\dagger\,\vac).
\label{2pointfunction}
\ee
This is given from eq.~(\ref{Wightmanxunibis}) by
\beaa{rcccl}
\wh{W}_o(y,y') &=& W_o(x(y),x(y')) &=&
- \sur{1}{4\pi}\ \log \left[\,x(y')-x(y)+\I 0^+ \,\right].
\label{eq:horizonGreen}
\eeaa

The energy-momentum observables in the incoming and outgoing 
coordinates are given by the products of derivatives of the field 
at the same point:
\be
\gT(x) &=& \partial_x \phi(x)^\dagger\,\partial_x \phi(x),
\label{TEIopx} \\ [2mm]
\wh{\gT}(y) &=& 
\partial_y \ch{\phi}(y)^\dagger\,\partial_y \ch{\phi}(y).
\label{TEIopy}
\ee
Their mean value in a given state must thus be regularized.
This regularization may be carried out in a covariant way along a 
geodesic by subtracting the mean value in Minkowski space-time
\cite{DF}, or by ordering the fields normally following
a covariant procedure \cite{MS}.
These two methods must give identical results and their 
application is made simpler in (asymptotically) flat space-time 
regions\footnote{The normal order regularization was
applied for the Dirac field in asymptotically flat space-time 
regions by Th.~Gallay and G.~Wanders 
\cite{GaW}.}.
The regularized mean value of $\wh{\gT}(y)$ in the incoming
vacuum will be computed here in the outgoing (asymptotically) flat 
space-time region $M_F$.
This is called the {\it energy-momentum tensor} and will be
denoted by $\wh{T}_o(y)$.

The observables $\gT_\gep (x)$ and $\wh{\gT}_\gep (y)$ are defined
by
\be
\gT_\gep (x) &=&\frac{1}{2} \left[ \,
\partial_x \phi(x)^\dagger \, \partial_x \phi(x+\gep) +
\partial_x \phi(x+\gep)^\dagger \, \partial_x \phi(x) \, \right], 
\label{TEIxbis}\\ [2mm]
\wh{\gT}_\gep (y) &=& \frac{1}{2} \left[ \,
\partial_y \ch{\phi}(y)^\dagger\,\partial_y \ch{\phi}(y+\gep) +
\partial_y \ch{\phi}(y+\gep)^\dagger\,
\partial_y \ch{\phi}(y)\,\right].
\label{TEIybis}
\ee
The energy-momentum tensor $\wh{T}_o(y)$ regularized by 
subtraction is given by the limit
\be
\wh{T}_o(y) &=& \lim_{\gep \rightarrow 0}\
(\vac, \left[ \,\wh{\gT}_\gep (y)\, -
\gT_\gep (x(y)) \, \right]\, \vac),
\label{defTEI} 
\ee
which is well defined.
It is computed using the representation (\ref{devphichapin}) or 
(\ref{fieldincomplex}) of the outgoing field in the 
incoming Hilbert space $\hin$ and is given by 
(see appendix \ref{ap:TEI})
\be
\wh{T}_o(y) &=& -\frac{1}{24 \pi} \, \SS{x(y)}{y},
\label{TEI}
\ee
where $\SS{x(y)}{y}$ is the Schwartzian derivative of $x(y)$ with 
respect to $y$
\footnote{
We have also
$\SS{x}{y}
= \sur{x'''}{x'}-\sur{3}{2}\, \left( \sur{x''}{x'} \right)^2 
= -2\,\sqrt{x'}\ \partial_y^2 \sur{1}{\sqrt{x'}}
= \partial_y^2 \log x' - \sur{1}{2} \, (\partial_y \log x')^2.$}:
\be
\SS{x}{y}=  \left( \sur{x''}{x'} \right)' 
-\sur{1}{2} \, \left( \sur{x''}{x'} \right)^2.
\ee
The energy-momentum tensor may also be regularized normally
as follows
\be
\wh{T}_o(y) &=& (\vac ,:\wh{\gT}\, (y) :_{out} \, \vac),
\label{defTEInormal}
\ee  
where the outgoing normal ordering has to be carried out before
computing the incoming vacuum mean value.
This definition also implies the result (\ref{TEI}) but in this 
case the computation is laborious 
(see appendix \ref{ap:TEInormal}).

From eq.~(\ref{TEI}) the transformation law for the 
energy-momentum tensor is deduced under the change of coordinates 
$y=y(z)$
\be
\wh{T}_o(y)\longrightarrow \wh{\wh{T}\,}\!\!_o(z)
&=& y'(z)^2 \ \wh{T}_o(y(z))-\frac{1}{24\pi}\,\SS{y(z)}{z},
\label{TEItransfo}
\ee
where $\wh{T}_o(y)$ and $\wh{\wh{T}\,}\!\!_o(z)$ are the 
regularized mean values of the energy-momentum observables 
in the incoming vacuum in the coordinates $y$ and $z$ 
respectively.

For the complex scalar field the incoming and outgoing current 
observables are given by
\be
\gU (x) &=& i \, \phi(x)^\dagger
\stackrel{\leftrightarrow}{\partial}_x \phi(x),
\label{courantx} \\ [2mm]
\wh{\gU} (y) &=& i \, \ch{\phi}(y)^\dagger
\stackrel{\leftrightarrow}{\partial}_y \ch{\phi}(y),
\label{couranty}
\ee
and the observables $\gU_\gep (x)$ and $\wh{\gU}_\gep (y)$ are
defined as
\be
\gU_\gep (x)&=& i\,\left[\
\phi(x+\gep)^\dagger \, \partial_x \phi(x)
- \partial_x \phi(x)^\dagger\, . \phi(x+\gep) \ \right],
\label{courantbisx} \\[2mm]
\wh{\gU}_\gep (y)&=& i\,\left[\
\ch{\phi}(y+\gep)^\dagger \, \partial_y \ch{\phi}(y)
- \partial_y \ch{\phi}(y)^\dagger\,. 
\ch{\phi}(y+\gep)\ \right].
\label{courantbisy}
\ee
The {\it outgoing current} $\wh{J}_o(y)$ is defined in the 
subtraction regularization scheme as
\be
\wh{J}_o(y) &=& \lim_{\gep \rightarrow 0} \ (\vac,
\left[ \, \wh{\gU}_\gep (y) -\gU_\gep (x(y))\,\right]
\, \vac ).
\label{defCOU}
\ee 
This limit is well defined and is computed in appendix 
\ref{ap:COU}\footnote{
The same result was obtained for the Dirac field \cite{GaW}.}:
\be
\wh{J}_o(y) &=& 0.
\label{COU}
\ee
The outgoing current vanishes for any transformation
of coordinates $x=x(y)$, i.e.~particles and antiparticles are
always created locally in pairs.

The outgoing current $\wh{J}_o(y)$ in the normal order
regularization scheme is defined by the equation
\be
\wh{J}_o(y) &=& (\vac, :\wh{\gU}(y):_{out} \, \vac ),
\label{defCOUnormal}
\ee
which also implies the result (\ref{COU})
(see appendix \ref{ap:COUnormal}).

In the real scalar fields,
the {\it mean number of spontaneously created particles} for a 
normalized particle test function $\tf{f}\in\ssp$ is defined by
\be
\bar{N}_o[f] &=& 
(\vac, a_{out}[\tf{f}]^\dagger \, a_{out}[\tf{f}]\,\vac),
\label{defN[f]}
\ee
and using the Bogoliubov transformations (\ref{bogoliubov}) this 
implies
\be
\bar{N}_o[f] &=& 
(\vac, a_{in}[B^*\tf{f}^*]\,a_{in}^{\dagger}[B\tf{f}]\,\vac).
\label{prenumber}
\ee
This quantity is thus expressed in terms of the Fourier 
transform $\tf{f}(p)$ by
\be
\bar{N}_o[f] &=& \no{B\tf{f}}^2,
\label{number}
\ee
showing that the mean number $\bar{N}_o[f]$ depends
only on the negative momentum contributions of the incoming
test function $\ch{f}(x)$.
$\bar{N}_o[f]$ is also expressed directly in 
terms of the outgoing test function $f(y)$ using 
eq.~(\ref{formulaeB})
\be
\bar{N}_o[f] &=& 
-\frac{1}{4\pi}\,\intii dy \intii dy'\, 
f(y)\,\log \left[\,\sur{x(y)-x(y')}{y-y'}\,\right]\,f(y')^*.
\label{numberbis} 
\ee
It may be checked that the l.h.s.~of eq.~(\ref{numberbis}) is 
always positive if $f(y)$ is a particle test function.
The results (\ref{number}) and (\ref{numberbis}) are extended
to any wave function $\tf{f}\in\LL$ if $f(y)$ exists a.e.~and is 
integrable.

The mean number of spontaneously created particles {\it in the 
mode} $f_p$, given by eqs (\ref{defmodep}) and (\ref{eq:series}), 
is defined formally as
\be
\bar{N}_o[f_p] &=& 4p^2\intoi \sur{dk}{2k}\,
\left\vert\,B(k,p)\,\right\vert^2,
\label{defmodeNp}
\ee
in agreement with eq.~(\ref{number}).
The {\it total mean number} $\bar{N}_o^{tot}$ of spontaneously 
created particles is defined as the sum of the contributions
(\ref{defmodeNp}) for each mode $f_p$,
\beaa{rcccl}
\bar{N}_o^{tot} &=& \intoi \sur{dp}{2p} \ \bar{N}_o[f_p]
&=& \intoi dp\,2p \intoi \sur{dk}{2k}\,
\left\vert\, B(k,p)\,\right\vert^2,
\label{Ntot} 
\eeaa
which can also be expressed as (see appendix 
\ref{ap:Ntotbis})\footnote{
Note that the kernel in the double integral (\ref{Ntotbis}) is not
symmetric as in the case of the Dirac field \cite{GaW}.}
\be
\bar{N}_o^{tot}  
&=& \sur{1}{4\,\pi^2}\,\intii dy\intii dy'\,
P\sur{1}{y-y'}\,
\left[\,\sur{x'(y)}{x(y)-x(y')}-\sur{1}{y-y'}\,\right].
\label{Ntotbis}
\ee

The operator $U$ is said to be {\it unitarily implementable} 
if there exists a unitary operator 
\mbox{$\cU:\hin \rightarrow \hout$} which satisfies
\be
\phi[f] &=& \cU^\dagger\ \ch{\phi}[f] \ \cU,
\saut \forall\,\tf{f}\in\so.
\ee 
If the operator $\cU$ exists, the fields $\phi$ and
$\ch{\phi}$ are equivalent representations of the commutator
(\ref{commutator}),
in the Hilbert spaces $\hin$ and $\hout$ respectively, and
the incoming and outgoing vacuums are related by
\be
\Psi_o &=& \cU \,\vac.
\label{vacuumsinout}
\ee
It has been proved that the operator $U$ is unitarily 
implementable if and only if $\bar{N}_o^{tot}$ is finite~\cite{Be}.

The definition (\ref{defN[f]}) of the mean number of spontaneously
created particles is generalized to an $n$-particle normalized 
test function $f^{(n)}$ by the equation
\be
\bar{N}_o[f^{(n)}] &=& (\vac, N_{out}[f^{(n)}] \, \vac ),
\label{defNfn}
\ee
where 
\be
N_{out}[f^{(n)}] &=& 
\ch{\phi}[f^{(n)}]^\dagger\,\ch{\phi}[f^{(n)}].
\label{defNfnbis}
\ee
Assuming that the one-particle test functions $f_i$ are
orthonormalized
\be
\bra \ind{\tf{f}}{i},\,\ind{\tf{f}}{j} \ket &=& \gd_{ij},
\label{orthonor}
\ee
eq.~(\ref{defNfn}) gives
\be
N_{out}[f^{(n)}] 
&=& N_{out}[f_1]\,N_{out}[f_2]\, ...\, N_{out}[f_n],
\ee
where the set of operators $N_{out}[f_i]$ satisfies
\be
\left[ \ N_{out}\,[f_i],\,N_{out}\,[f_j] \ \right] &=& 0, \saut
i,j= 1,2,...,n,
\ee
and where $f^{(n)}$ is defined by eqs (\ref{deffbar}) and 
(\ref{fbarnormal}).
Under the assumption (\ref{orthonor}) it is possible to give a 
compact formula for the mean number $\bar{N}_o[f^{(n)}]$
using the definitions
\be
\mbox{}\hspace{0mm}
( f^{(n)}\times {f^{(n)}}^*)_{_S} \,(y_1,...,y_{2n}) 
= \sur{C^2}{(2n)!}\,\SSum_{\tau\in {\cal P}_{2n}}
f_1(y_{\tau(1)})  \, ... \,
f_n(y_{\tau(n)}) \, f_1(y_{\tau(n+1)})^* \, 
 ... \, f_n(y_{\tau(2n)})^* \hspace{5mm}
\label{symetrisation}
\ee
and
$G^n=\overbrace{G \times G \times ...\times G}^{n\ {\rm times}}$,
where $G$ is defined by eq.~(\ref{defG}).
This formula is displayed in the following theorem, proved
in appendix \ref{ap:Nnparticles}.
\begin{theorem} \label{th:Nnparticles}
If $\{ \ind{\tf{f}}{i} \}_{i=1}^n \subset \LL$ is an orthonormal
set of functions such that $f_i$ exists and is integrable
($i=1,2,...,n$), then
\be 
\bar{N}_o[f^{(n)}] &=& C^2\ \sur{(2n)!}{2^n\,n!} \ G^n \, 
[ \, (f^{(n)}\times {f^{(n)}}^*)_{_S} \, ],
\label{Nnparticles}
\ee
where $f^{(n)}$ is defined by eqs (\ref{deffbar}) and 
(\ref{fbarnormal}).
Equation (\ref{Nnparticles}) contains at most 
$\frac{2^n\,n!}{(2n)!}$ distinct terms.
\end{theorem}
\section{Scalar field theory in a thermal bath}
\label{sec:thermal}
In this section the one-dimensional massless scalar field 
is considered in a thermal bath of temperature $\gb^{-1}$ for null 
chemical potential.
I will restrict the scalar field to the finite interval $[-L,L]$
and impose periodic boundary conditions before taking the 
``thermodynamic" limit $L\rightarrow \infty$.
This procedure is necessary to define thermal mean values correctly.
The space-time and energy-momentum variables will be denoted here 
by $\tau$ and $\go$.

The real scalar field $\gp_{_L}(\tau)$ in the interval $[-L,L]$ is 
given by 
\be
\gp_{_L} (\tau) &=& \frac{1}{\sqrt{2L}} \sum_{n=1}^\infty
\sur{1}{2\go_n} \left[ \, a_n \, e^{-\I\go_n \tau} 
+ a_n^\dagger \,e^{\I\go_n \tau} \, \right] 
+ \frac{a_o}{\sqrt{2L}},
\label{fieldL}
\ee
where $\tau\in [-L,L]$, $\go_n=n\pi /L$, $a_o\in\R$ 
and $a_n\in\C$ if $n\in\N$.
It is quantized by imposing the field operator 
commutators
\beaa{rcccccl}
\left[\, a_n,\ a_m^\dagger\,\right] &=& 2\,\go_n\,\gd_{n,m},
&\saut&
\left[\, a_n,\ a_m\,\right] &=& 0,
\label{Thermalcommutators}
\eeaa
where $n,m \geq 1$.
These act on a Hilbert space $\hh_{_L}$ whose vacuum will be 
denoted by $\Phi_o$.
Equations (\ref{fieldL}) and (\ref{Thermalcommutators}) show
that the field commutator is given by\footnote{Using the formula
$\sum_{n=1}^\infty \sur{1}{n}\, \sin (a\,n)
= \signe{a}\ \sur{\pi-\abs{a}}{2}$ \ ($\abs{a}\ <2\pi$),
with $a=\pi\,(\tau'-\tau)/L$ \cite{Ha}.}
\be
[\, \gp_{_L}(\tau), \gp_{_L}(\tau')\, ] 
&=& \sur{\I}{4}\, \signe{\tau'-\tau}\, 
\left(\, 1-\sur{\abs{\tau'-\tau}}{L} \,\right),
\label{thermalcomfield} 
\ee
where $\tau,\tau'\in [-L,L]$.

The correspondence between field theories for finite and infinite
intervals is given by
\beaa{rcl}
R_{_L}, \,\cH_{_L},\,\gp_{_L} & \longleftrightarrow  & 
R_\infty, \,\cH,\,\gp,
\\ [1.5mm]
\go_n, \ n\in\N & \longleftrightarrow  & \go\in\R_+^*, 
\\ [1.5mm]
L\,\gd_{n,m} & \longleftrightarrow & \pi\,\gd(\go-\go'), 
\\ [1.5mm]
\sqrt{\pi}\, a_n & \longleftrightarrow & \sqrt{L}\, a(\go).
\eeaa
In the thermodynamic limit the null momentum mode $a_o$ 
disappears in eq.~(\ref{fieldL}) and the commutator 
(\ref{thermalcomfield}) is then formally equal to 
eq.~(\ref{commutator2}).

The {\it thermal mean value} of a given observable $A$ is defined by 
the limit
\be
\mt{A} &=& \lim_{L\rightarrow\infty} \sur{ 
\Tr_{_L} \left[ \, e^{-\gb H_{_L}} \, A \, \right]}
{\Tr_{_L} \left[ \, e^{-\gb H_{_L}} \, \right]},
\label{moyennethermique}
\ee
where $\Tr_{_L}$ is the trace on the Hilbert space $\cH_{_L}$ and
$H_{_L}$ is the free Hamiltonian given by
\be
H_{_L} &=& \sum_{n=1}^\infty \go_n\,a_n^\dagger\,a_n.
\ee
The partition function 
$Z_{_L}=\Tr_{_L} \left[ \, e^{-\gb H_{_L}} \, \right]$ is IR 
divergent in the thermodynamic limit.
Note that the thermal mean value (\ref{moyennethermique}) is 
generally well defined although this limit will not necessarily 
converge to a finite value for any observable $A$.

From def.~(\ref{moyennethermique}), thermal mean values of field 
operators are given by 
\be
\mt{a^\dagger(\go)\, a(\go')} &=&   
\sur{2\go}{e^{\gb \go}-1}\ \gd(\go-\go'),
\label{thermalfieldop} \\
\mt{ a(\go)\, a^\dagger(\go')} &=&
\sur{2\go}{1-e^{-\gb\go}} \ \gd(\go-\go'),
\label{thermalfieldopbis} \\ [2mm]
\mt{a(\go)\, a(\go')} &=& 
\mt{a^\dagger(\go)\, a^\dagger(\go')} \ = \ \,0,
\label{thermalfieldopter}
\ee
where $\go,\go'>0$.
Equation (\ref{thermalfieldop}) is proved in appendix 
\ref{ap:thermalfieldop}.
We also have
\beaa{l}
\mt{a^\dagger(\go_n)\, ... \, a^\dagger(\go_1)
\, a(\go_1') \ ... \ a(\go_n')}
\hspace{80mm} \\ [3mm]\hspace{30mm} = \ \
\sur{2\go_1}{e^{\gb\go_1}-1} \ {\cdot\cdot\cdot} \
\sur{2\go_n}{e^{\gb\go_n}-1} \, 
\SSum_{\gs\in {\cal P}_n}
\gd(\go_1-\go_{\gs(1)}') \ \, ... \, \
\gd(\go_n-\go_{\gs(n)}'),
\label{thermalfieldopn}
\eeaa
where $\go_i,\go_i'>0, \,i=1,2,...,n$ (see appendix 
\ref{ap:thermalfieldopn}).
More generally, the Wick theorem is satisfied for thermal mean
values of field operators.

The {\it thermal two-point function} is defined by
\be
W_\gb^{Th}\, (\tau,\tau') &=& \mt{\phi(\tau)\,\phi(\tau')^\dagger},
\label{defthermal2pf}
\ee 
and satisfies the properties
\be
W_\gb^{Th}\, (\tau,\tau') &=& W_\gb^{Th}\, (\tau,\tau'+\I n\gb),
\saut \forall\,n\in\Z,
\label{periodic2pfthermal} \\ [2mm]
\Re W_\gb^{Th}\, (\tau,\tau') 
&=& \sum_{n=-\infty}^{+\infty} 
\Re W^{Th}_{\infty} \, (\tau+\I n\gb,\tau'),
\label{thermalsomme}
\ee
$\forall\, \tau,\tau'\in\R$ (see appendix \ref{ap:thermalsomme}).
Using the formula \cite[(89.10.4)]{Ha}
\be
(\tau'-\tau)^2 
\Prod_{n=1}^\infty \sur{(\tau'-\tau+\I n\gb)^2 \, 
(\tau'-\tau-\I n\gb)^2}{\gb^4 n^4} &=& 
\sur{\gb^2}{\pi^2}\ \sinh^2 
\left[\,\sur{\pi}{\gb}\,(\tau'-\tau)\,\right],
\ee
we obtain from eqs (\ref{Wightmanxunibis}) and (\ref{thermalsomme})
\be
\Re W_\gb^{Th} \, (\tau,\tau') &=& -\sur{1}{4\pi}
\log \left[\,\frac{\gb}{\pi} 
\sinh \left(\frac{\pi}{\gb} \abs{\tau'-\tau}\right) \right]+ C,
\label{Rethermal}
\ee
where $C$ is an infinite constant, hence the thermal mean 
value (\ref{defthermal2pf}) is infinite.

The thermal two-point function $W_\gb^{Th}(\tau,\tau')$ will thus be 
redefined as the kernel of a distribution on 
$\so\times\so$ by
\be
W^{Th}_\gb[\,f_1\times f_2^*\,] &=&
\mt{ \gp[f_1] \, \gp[f_2]^\dagger}.
\label{defthermal2pfbis}
\ee
From eqs (\ref{fieldL}) and (\ref{thermalfieldop}) to 
(\ref{thermalfieldopter}) we obtain
\be
W^{Th}_\gb[\,f_1\times f_2^*\,] &=&
\intii \sur{d\go}{2\go} \, 
\sur{\ind{\tf{f}}{2}(\go)^*\,\ind{\tf{f}}{1}(\go)}{1-e^{-\gb\go}},
\label{defthermal2pfter}
\ee
where $\ind{\tf{f}}{1},\ind{\tf{f}}{2}\in\so$.
The following theorem, proved in appendix \ref{ap:2pfthermal},
gives the correct expression for the kernel 
$W_\gb^{Th}\, (\tau,\tau')$. 
\begin{theorem}
Between kernels of distributions on $\so\times\so$,
\be
W_\gb^{Th}\, (\tau,\tau') &=&
- \sur{1}{4\pi} \log \left\{\, \sur{\gb}{\pi} \,
\sinh \left[\,\frac{\pi}{\gb}\,\left(\tau'-\tau+\I 0^+\right) 
\,\right]\,\right\}.
\label{2pfthermal}
\ee
\end{theorem}
The periodicity property (\ref{periodic2pfthermal}) is satisfied 
by (\ref{2pfthermal}) up to an irrelevant constant.

The {\it thermal energy-momentum tensor} $T_\gb^{Th}(\tau)$ is 
defined by the limit
\be
T_\gb^{Th} (\tau) &=& \lim_{\gep\rightarrow 0}
\left[\,  \mt{\gT_\gep (\tau)}- (\Phi_o,\gT_\gep (\tau)\,\Phi_o)
\,\right],
\ee
where the observable $\gT_\gep(\tau)$ is given by
\be
\gT_\gep (\tau) &=& \frac{1}{2} \left[ \,
\partial_\tau \gp(\tau)^\dagger\,\partial_\tau \gp(\tau+\gep) +
\partial_\tau \gp(\tau+\gep)^\dagger\,
\partial_\tau \gp(\tau)\,\right].
\ee
Using eq.~(\ref{2pfthermal}) we obtain
\beaa{rcccl}
\mt{\gT_\gep (\tau)} &=&
-\sur{\pi}{4\gb^2} \, 
\sur{1}{\sinh^2 \left(\pi\,\gep/\gb \right)}  &=&
-\sur{1}{4\pi^2 \gep^2}+ \sur{\pi}{12 \gb^2} + {\cal O}(\gep^2),
\eeaa
from which we deduce that $T_\gb^{Th}(\tau)$ depends only on $\gb$:
\beaa{rcccl}
T_\gb^{Th}(\tau) &=& T_\gb^{Th}
&=& \sur{\pi}{12 \gb^2}, \saut \forall\, \tau\in\R.
\label{TEIthermal}
\eeaa

The {\it thermal current} $J^{Th}_\gb(\tau)$ associated with the 
complex scalar field is defined by
\be
J^{Th}_\gb(\tau) &=& \lim_{\gep \rightarrow 0}
\left[\ \mt{\gU_\gep(\tau)} -
(\Phi_o, \gU_\gep (\tau)\,\Phi_o)\ \right],
\label{thermaldefCOU}
\ee 
where the observable $\gU_\gep(\tau)$ is given by
\be
\gU_\gep (\tau) &=& i \left[\,
\gp(\tau+\gep)^\dagger \, \partial_\tau \phi(\tau)
- \partial_\tau \gp(\tau)^\dagger\, . \gp(\tau+\gep) \, \right].
\label{currentthermalbis}
\ee
The limit (\ref{thermaldefCOU}) is well defined and is given by
\be
J^{Th}_\gb(\tau) &=& 0,
\label{thermalCOU}
\ee
so there is no net local current.

In the real scalar fields,
the {\it thermal mean value of the number of particles} for a 
normalized particle test function \mbox{$f\in\ssp$} is defined by
\be
\bar{N}_\gb^{Th}[f] &=& \mt{ a[f]^\dagger \, a[f]},
\label{th:thermiqueNDEF}
\ee
and from eq.~(\ref{thermalfieldop}) we obtain
\be
\bar{N}_\gb^{Th}[f]
&=& \int_0^\infty \sur{d\go}{2\go} \sur{\abs{\tf{f}(\go)}^2}
{e^{\gb\go}-1}\cdot
\label{Nthermal}
\ee
This result is extended to any wave function $\tf{f}\in\LLw$ if 
$f(\tau)$ exists a.e.~and is integrable.

We define furthermore the distribution $G_\gb^{Th}$ on 
$\LLw\times\LLw$ by
\be
G_\gb^{Th} (f_1 \times f_2^*) &=& 
\int_0^\infty \sur{d\go}{2\go} 
\sur{\ind{\tf{f}}{2}(\go)^*\ind{\tf{f}}{1}(\go)}{e^{\gb\go}-1}
\cdot
\label{Gthermal}
\ee
The following theorem gives an expression for the thermal mean value
of the number of particles for an $n$-particle normalized test 
function $f^{(n)}$:
\be
\bar{N}^{Th}_\gb[f^{(n)}] &=& 
\mtout{\ch{\phi}[f^{(n)}]^\dagger\,\ch{\phi}[f^{(n)}]}.
\label{defNfnthermal}
\ee
It is easily proved using eq.~(\ref{thermalfieldopn}).
\begin{theorem} \label{th:thermiqueNgeneral}
If $\{ \ind{\tf{f}}{i} \}_{i=1}^n \subset \LLw$ is an orthonormal
set of functions such that $f_i$ exists and is integrable
($i=1,2,...,n$), then
\be
\bar{N}^{Th}_\gb[f^{(n)}] &=&  
C^2\SSum_{\gs \in {\cal P}_n}\ 
G_\gb^{Th}(f_1 \times f_{\gs(1)}^*) \ 
G_\gb^{Th} (f_2 \times f_{\gs(2)}^*) \ ...\ 
G_\gb^{Th} (f_n \times f_{\gs(n)}^*),\ \ \
\label{eq:thermiqueNgeneral}
\ee
where $f^{(n)}$ is defined as in eqs (\ref{deffbar}) and 
(\ref{fbarnormal}).
\end{theorem}

A state $\gP \in \hh$ is said to be a {\it thermal state} of 
temperature $\gb^{-1}$ if it satisfies the equation~\cite{FR}
\be
(\gP, A_\tau \ B\, \gP) &=& (\gP, B\, A_{\tau+\I\gb}\,\gP),
\label{KMS}
\ee
where $A$ and $B$ are two operators and where we have defined
\be
A_\tau &=& e^{\I\tau H} \, A \, e^{-\I\tau H},
\ee
where $H$ is the free Hamiltonian.
Equation (\ref{KMS}) is known as the KMS condition. 
It can also be written in the equivalent form \cite{HNS}
\be
(\gP,A\, B\, \gP) &=& \frac{1}{2\pi}\, \intii d\go \intii d\tau \,
\sur{e^{\I\go \tau}}{1-e^{-\gb\go}} \ (\gP,[\,A_\tau, B\,]\, \gP).
\label{preKMS}
\ee
In the particular case where $A=\gp(\tau)$ and $B=\gp(\tau')$, 
we obtain
\be
(\gP, \gp(\tau)\, \gp(\tau')\, \gP)
&=& \sur{1}{2\pi}\,\intii \frac{d\go}{2\go} \, 
\sur{e^{\I\go\,(\tau'-\tau)}}{1-e^{-\gb\go}}
\label{KMSbis}
\ee
from the commutator (\ref{thermalcomfield}).
The integral in the r.h.s.~is IR divergent and is formally equal to 
the kernel of \mbox{$W_\gb^{Th}[\,f_1\times f_2^*\,]$}
(see eq.~(\ref{defthermal2pfter})).
The KMS condition is thus restated as an equality between kernels of
distributions on $\so\times\so$ in the form
\be
(\gP,\gp(\tau)\, \gp(\tau')\, \gP) &=& W_{\gb}^{Th}(\tau,\tau'),
\label{KMSfield}
\ee
where $W_{\gb}^{Th}(\tau,\tau')$ is given by eq.~(\ref{2pfthermal}).
If this last equation is satisfied on a interval $I$ for a given 
state $\gP$, $\forall\,\tau,\tau'\in I$, we shall say that $\gP$ is
a thermal state on this interval.
\section{Spontaneous creation of particles}
So far the massless scalar field has been studied in 2D curved 
space-times.
In this section the results obtained previously are applied to the 
relativistic black hole model, for which the transformation of 
coordinates $x=x(y)$ is given by (\ref{transfo})
\be
x(y) &=& \gD- e^{-My},\saut\forall\,y\in\R.
\label{transfoTNR}
\ee

The kernel $U(k,p)$, defined by eq.~(\ref{KernelU}), can be explicitly 
computed for this model and is given by (see appendix \ref{ap:UnoyauTNR})
\be
U(k,p) &=& 
\sur{e^{-\I k\gD}\,e^{-\I \gO\left(\frac{p}{M}\right)}\,
e^{\I \frac{p}{M} \log \vert k \vert}}{\sqrt{2\pi\,M}}\ 
\left[ \ \sur{\gt(k)}{\sqrt{p\,(1-e^{-\frac{2\pi}{M} p})}}
+\sur{\gt(-k)}{\sqrt{p\,(e^{\frac{2\pi}{M}p}-1)}} \ \right],
\label{UkernelTNR}
\ee
$\forall\,k,p\not= 0$, where $\gO(p)={\rm Arg}\,[\gC(\I p)]$.
Note that this kernel satisfies the property
\be
\left\vert\,U(k,p)\,\right\vert &=& 
e^{\sgn(k) \pi p}\,\left\vert\,U(-k,p)\,\right\vert.
\ee

The Bogoliubov transformation (\ref{Bogoliubovbis}) is obtained from 
eq.~(\ref{UkernelTNR}) and is given by
\be
a_{out}(p) &=& \sqrt{\sur{Mp}{2\pi}} \ 
e^{-\I k\gD}\,e^{-\I\gO\left(\frac{p}{M}\right)}
\int_0^\infty \sur{dk}{k} \, e^{\I\frac{p}{M}\log k} \left[ \
\sur{a_{in}(k)}{\sqrt{1-e^{-\frac{2\pi}{M}p}}} 
+ \sur{a_{in}^\dagger(k)}{\sqrt{e^{\frac{2\pi}{M}p}-1}} \ \right], 
\label{eq:horizonbogoliubov}
\ee
where $p>0$.
The kernel (\ref{UkernelTNR}) and the Bogoliubov transformation
(\ref{eq:horizonbogoliubov}) are not invertible (see discussion following
eq.~(\ref{eq:applicationVC})).

Equations (\ref{defmodeNp}) and (\ref{UkernelTNR}) show that the 
mean number of spontaneously created particles for the mode $f_p$ 
(\ref{defmodep}) is IR and UV divergent in the incoming momentum $k$:
\beaa{rcccl}
\bar{N}_o[f_p] &=& \sur{1}{\pi M}\ \sur{2p}{e^{\frac{2\pi}{M} p}-1}
\intoi \sur{dk}{2k} &=& \infty,
\eeaa
if $p>0$.
This result is also true for $p=0$ in which case $f_{_0}$ is given by
def.~(\ref{eq:series}).
The total mean number of spontaneously created particles is moreover IR 
divergent in the outgoing momentum $p$ (see eq.~(\ref{Ntot}))
\be
\bar{N}_o^{tot} &=& \infty,
\label{Ntotal}
\ee
and the operator $U$ is therefore not implementable
(see discussion after eq.~(\ref{vacuumsinout})).

In the following, the mean values of outgoing observables in the incoming 
vacuum are compared with their corresponding thermal mean values in the 
Hilbert space $\hout$, given by (see eq.~(\ref{moyennethermique}))
\be
\mtout{A} &=&  \lim_{L\rightarrow\infty} \sur{
{\Trout}_{^L}\left[\,e^{-\gb H_{L,out}}\,A\,\right]}
{{\Trout}_{^L}\left[\,e^{-\gb H_{L,out}}\,\right]}\cdot
\ee
This enables us to establish the thermal properties of the radiation 
emitted, and in particular to determine its temperature.

The outgoing two-point function (\ref{2pointfunction}) is given for
the transformation (\ref{transfoTNR}) by
\be
\wh{W}_o \, (y,y') &=& -\sur{1}{4\pi} \, 
\log\left( \,e^{-My}-e^{-My'}+\I 0^+\,\right).
\label{2pfuntionTNR}
\ee
Writing the thermal two-point function (\ref{2pfthermal}) in the form
\be
W_{\gb,out}^{Th}\, (y,y') &=& -\frac{1}{4\gb} \,(y+y')
- \frac{1}{4\pi} \log \left[\, \frac{\gb}{2\pi} \,
\left(\, e^{-\frac{2\pi}{\gb} y} 
- e^{-\frac{2\pi}{\gb} y'}+\I 0^+\, \right) \, \right],
\label{2pfuntionTh}
\ee 
we deduce that the two-point functions (\ref{2pfuntionTNR}) and 
(\ref{2pfuntionTh}) are equivalent everywhere as kernels of 
distributions on $\so\times\so$, if and only if $\gb = \frac{2\pi}{M}$:
\be
\wh{W}_o\,(y,y') &=& W_{\frac{2\pi}{M},out}^{Th}\,(y,y'), 
\saut \forall\,y,y'\in\R.
\label{KMSTNR}
\ee
We conclude from this last equation that the incoming vacuum $\vac$ is 
a thermal state of temperature $\frac{M}{2\pi}$ in the 
outgoing coordinates on $\R$.

The energy-momentum tensor is computed from eq.~(\ref{TEI})
and is given 
by
\be
\wh{T}_o(y) &=& \sur{M^2}{48\pi}, \saut \forall\,y\in\R;
\ee
hence we deduce from eq.~(\ref{TEIthermal}) that it is thermal
\be
\wh{T}_o(y) &=& T_{\frac{2\pi}{M},out}^{Th}, \saut \forall\,y\in\R,
\ee
and that the associated temperature is also given by $\frac{M}{2\pi}$ 
for all $y\in\R$. 

We consider now the mean number of spontaneously created particles
for a given normalized particle function $f$.
If $f$ is a Schwartz function, eq.~(\ref{numberbis}) shows that
$\bar{N}_o[f]$ is always finite for the transformation 
(\ref{transfoTNR}):
\be
\bar{N}_o[f] &<& \infty, \saut \forall\,\tf{f}\in\ssp.
\label{Nfinite}
\ee
The mean number of particles $\bar{N}_o[f]$  may be explicitly computed 
from eq.~(\ref{number}) and (\ref{UkernelTNR}) and is given by 
(see appendix \ref{ap:NTNR})
\be
\bar{N}_o[f] &=&
\int_0^\infty \sur{dp}{2p} \ 
\sur{\abs{\tf{f}(p)}^2}{e^{\frac{2\pi}{M}p}-1}\cdot
\label{NTNR}
\ee
This result shows that $\bar{N}_o[f]$ may also be infinite.
For example, defining the test functions $\ind{\tf{f}}{\ga}\in\LL$ by
\be
\ind{\tf{f}}{\ga}(p) &=& C_\ga\,\gt(p)\,p^\ga\,e^{-p^2},
\saut \ga>0,
\ee
where $C_\ga$ is a normalization constant, we have the 
equivalence
\be
\bar{N}_o[f_\ga] \ = \ \infty &\Longleftrightarrow& \ga\leq 1/2. 
\ee
If $\ga\leq 1/2$, $\bar{N}_o[f_\ga]$ is IR divergent in the outgoing 
momentum $p$.

Comparing eq.~(\ref{NTNR}) with the thermal expression 
(\ref{Nthermal}), we deduce that the mean number of spontaneously 
created particles is thermal
\be  
\bar{N}_o[f] &=& \bar{N}_{\frac{2\pi}{M},out}^{Th}\,[f],
\label{NTNRthermal}
\ee 
and that the associated temperature is also given by $\frac{M}{2\pi}$.
This last result is also true for a normalized $n$-particle test function 
$f^{(n)}$.
This can easily be proved (see appendix \ref{ap:Ngeneral}) in the special
case for which the functions $f_i$ are orthonormalized, as stated in the 
following theorem.
\begin{theorem} \label{th:Ngeneral}
If $\{ \ind{\tf{f}}{i} \}_{i=1}^n \subset \LL$ is a set of normalized 
test functions such that $f_i$ exists and is integrable ($i=1,2,...,n$),
then
\be
\bar{N}_o[f^{(n)}] &=& \bar{N}^{Th}_{\frac{2\pi}{M},out}[f^{(n)}],
\label{Ngeneral}
\ee 
where $f^{(n)}$ is defined by eqs (\ref{deffbar}) 
and (\ref{fbarnormal}).
\end{theorem}
\section{Conclusions}
This new space-time model, based on the ``$R=T$" relativistic 
theory, describes the formation of a black hole  
whose semi-classical approach is straightforward.

This black hole emits an infinity of massless particles in each
outgoing momentum mode.

The emission is thermal in the sense that mean values in the 
incoming vacuum of observables constructed in the outgoing 
coordinates  are equal to their thermal averages:
\be
(\vac, \,A\,\vac) &=& \mtbout{A}{\frac{2\pi}{M}}.
\label{thermalbehaviour}
\ee
Immediately after the formation of the black hole
this result is valid everywhere, and not only near the horizon.
%
Equation (\ref{thermalbehaviour}) shows that the temperature 
of the radiation is given by
\be
T_{\mbox{\tiny radiation}} &=& \sur{M}{2\pi},
\ee
and it is proportional to the relative amplitude of the 
localized curvature (\ref{modeldef}).
The radiation emitted by the black hole is thus described by an 
outgoing density matrix which is thermal.
\\ \\
{\large \bf Acknowledgments} 
\\ \\
I thank G.~Wanders for stimulating discussion and criticism,
and D.~Rickebusch for correcting the manuscript.
\appendix\section{Appendices}
If $f$ is an integrable function, the primitives $F(y)$ and 
$\cha{F}(x)$ are defined as
\beaa{rclcrcl}
F(y) &=& \Int_{-\infty}^y dy'\,f(y'), &\saut&
\cha{F}(x) &=& \Int_{x(-\infty)}^x dx'\,\ch{f}(x').
\label{apformulaeA}
\eeaa
They are related by  
\be
\cha{F}(x(y)) &=& F(y), \saut \forall\,y\in\R,
\label{FTransformation}
\ee 
and satisfy
\beaa{rclcrcl}
\tfa{F}(p) &=& \I p\,\tf{f}(p), &\saut& 
\tfcha{F}(k) &=& \I k\,\tfch{f}(k),
\label{apformulaeC}
\eeaa
\mbox{}\vspace{-13mm}
\\
\beaa{rcccccccl}
F(-\infty) &=& F(+\infty) &=& \cha{F}(x(-\infty))
&=& \cha{F}(x(+\infty)) &=& 0,
\label{apformulaeB}
\eeaa
if $\tf{f}(0)=0$.
\sub{Proof of eq.~(\ref{formulaeB})} 
\label{ap:formulaeB}
Definitions (\ref{apformulaeA}) show that
\beaa{rcccl}
\bra B\ind{\tf{f}}{2},B\ind{\tf{f}}{1} \ket
&=& - \sur{\I}{2}\,\intoi dk\,\indice{\tfch{f}}{2}(-k)^*\,
\indice{\tfcha{F}}{1}(-k) 
&=& \sur{1}{4\pi}\,\Int_I dx\,\indice{\ch{f}}{2}(x)^* 
\Int_I dx'\sur{\indice{\cha{F}}{1}(x')}{x'-x+\I 0^+},
\label{apformulaeD}
\eeaa
where $I=\{\,x(y)\, \mid\, y\in\R \, \}$.
Integrating by parts we obtain
\be
\sur{1}{4\pi}\int_I dx'\, \sur{\indice{\cha{F}}{1}(x')}{x'-x+\I 0^+} 
&=&
- \sur{1}{4\pi}\intii dy' \, f_1(y')\, \log \abs{x(y')-x(y)} 
- \sur{\I}{4}\,\indice{\cha{F}}{1}(x).
\label{apformulaeE}
\ee
The transformations (\ref{fTransformation}) and (\ref{FTransformation})
imply
\beaa{rclrcl}
-\sur{\I}{4}\,\Int_I dx\,
\indice{\ch{f}}{2}(x)^*\,\indice{\cha{F}}{1}(x) 
&=& -\sur{\I}{4}\,\intii dy \, f_2(y)^* F_1(y) 
&=& -\sur{1}{2} \, \intoi \sur{dp}{2p} \, 
\ind{\tf{f}}{2}(p)^* \,\ind{\tf{f}}{1}(p),
\label{apformulaeF}
\eeaa
and from eqs (\ref{apformulaeD}) to (\ref{apformulaeF}) 
\beaa{rcl}
\bra B\ind{\tf{f}}{2},B\ind{\tf{f}}{1} \ket  &=&
- \sur{1}{4\pi}\intii dy \intii dy' \, 
f_1(y')\, \log \abs{x(y')-x(y)}\,f_2(y)^* \\ [3mm]
&& -\sur{1}{2} \, \intoi \sur{dp}{2p} \, 
\ind{\tf{f}}{2}(p)^* \,\ind{\tf{f}}{1}(p).
\label{apformulaeG}
\eeaa
Restricting eq.~(\ref{apformulaeG}) to the identity transformation 
$x(y)=y$, we obtain
\be
\intoi \sur{dp}{2p} \,\ind{\tf{f}}{2}(p)^* \,\ind{\tf{f}}{1}(p) 
&=& - \sur{1}{2\pi}\intii dy \intii dy'\, 
f_1(y')\, \log \abs{y'-y}\,f_2(y)^*,
\label{apformulaeH}
\ee
and hence the result (\ref{formulaeB}) from eq.~(\ref{apformulaeG}).
\sub{Proof of eq.~(\ref{commutatorequality})}
\label{ap:commutatorequality}
The definitions (\ref{apformulaeA}) and transformations 
(\ref{fTransformation}) and (\ref{FTransformation}) show that
\beaa{rclrcl}
\intii \sur{dk}{2k} \, \ind{\tfch{f}}{2}(k)^* 
\,\ind{\tfch{f}}{1}(k) 
&=& \sur{\I}{2}\,\intii dk\,\ind{\tfch{f}}{2}(k)^*\,\tfcha{F}_1(k) 
&=& \sur{\I}{2}\,\intii dx \,\indice{\ch{f}}{2}(x)^*\,\cha{F}_1(x) 
\nonumber \\ [3mm]
&=& \sur{\I}{2}\,\intii dy \, f_2(y)^* \, F_1(y) 
&=&  \intii \sur{dp}{2p} \, \ind{\tf{f}}{2}(p)^* \,
\ind{\tf{f}}{1}(p).
\eeaa
\sub{First proof of eq.~(\ref{TEI}) } 
\label{ap:TEI}
The energy-momentum tensor is computed here from definition 
(\ref{defTEI}).
From the field representation (\ref{devphichapin}) or 
(\ref{fieldincomplex}) of the field $\ch{\phi}(y)$ in the Hilbert space 
$\hin$ we deduce
\be
(\vac,
\partial_y \ch{\phi}(y)^\dagger \, 
\partial_y \ch{\phi}(y+\gep) \, \vac) &=&
\sur{x'(y) \, x'(y+\gep)}{4\pi}\, 
\intoi dk \ k \ e^{\I k\,[x(y+\gep)-x(y)]}.
\label{AformuleTEI}
\ee
The formula
\be
\intoi dk \, k \ e^{\I kx} &=& -\sur{1}{(x+\I 0^+)^2},
\ee
and eq.~(\ref{AformuleTEI}) show that
\be
\wh{T}_o(y) &=& 
-\sur{1}{4\pi} \, \lim_{\gep \rightarrow 0}
\,\left\{ \, 
\sur{x'(y+\gep)\,x'(y)}{\left[\,x(y+\gep)-x(y)\,\right]^2}
- \sur{1}{\gep^2} \,\right\}.
\label{ATEI}
\ee
This limit is well defined and by expanding at $\gep=0$ we obtain
(\ref{TEI}).
\sub{Second proof of eq.~(\ref{TEI})} 
\label{ap:TEInormal}
The energy-momentum tensor is computed here from definition 
(\ref{defTEInormal}) and for the real scalar field.
Ordering normally the field operators and using the equations
\be
(\vac, a_{out}(p) \, a_{out}(p') \, \vac)
&=& 4 \,p\, p' \intoi \sur{dk}{2k} \, A(k,p) \,B(k,p'), 
\label{ATEIvma} \\ [2mm]
(\vac, a_{out}(p)^\dagger \, a_{out}(p') \, \vac)
&=& 4 \,p\, p' \intoi \sur{dk}{2k} \, B(k,p)^* \,B(k,p'),
\label{ATEIvmb}
\ee
deduced from the Bogoliubov transformations (\ref{bogoliubov}),
and then integrating on the momentum variables, 
we obtain
\be
(\vac, : \wh{\gT} (y):_{out} \, \vac) 
&=& -\sur{1}{16\pi^3}\, \intii dy' \intii dy'' \,
\sur{1}{\left[ \,x(y')-x(y'')-\I 0^+ \, \right]^2}\ \times 
\hspace{20mm}
\ee
\be
\nonumber \\ [-6mm] \hspace{-1.55mm} \nonumber
\left[\sur{1}{(y-y'+\I 0^+)(y-y''+\I 0^+)}
+ \sur{1}{(y-y'-\I 0^+)(y-y''-\I 0^+)} 
- \sur{2}{(y-y'+\I 0^+)(y-y''-\I 0^+)}\right].
\ee
Using\footnote{We have defined
$P \sur{1}{x^m} = \frac{(-1)^{m-1}}{2\,(m-1)!}
\Lim_{\gep \rightarrow 0} \sur{d^m}{dx^m} \,\log(x^2+\gep^2)$.}
\beaa{rcccccl}
\sur{1}{y \pm \I 0^+} &=& P\sur{1}{y} \,\mp\,\I\pi\,\gd(y), &\saut&
\sur{1}{\left( \,x\pm\I 0^+ \right)^2}
&=& P\sur{1}{x^2} \,\pm\,\I\pi\,\gd'(x),
\eeaa
we deduce the result
\beaa{l}
\mbox{} \hspace{-3mm}
\wh{T}_o(y) \ = \ 
-\sur{1}{4\pi} \intii dy' \intii dy'' \,\gd(y-y') \, \gd(y-y'')
\left\{ \sur{x'(y')\,x'(y'')}
{\left[\,x(y')-x(y'')\,\right]^2}-\sur{1}{(y'-y'')^2}\right\} \ \
\eeaa
which again gives the limit (\ref{ATEI}).
\sub{First proof of eq.~(\ref{COU})} \label{ap:COU}
The outgoing current is computed from def.~(\ref{defCOU}).
From the field representation (\ref{fieldincomplex}) of the
field $\ch{\phi}(y)$ in the Hilbert space $\hin$ we deduce
\be
(\vac, \ch{\phi}(y+\gep)^\dagger \, \partial_y \ch{\phi}(y) \,
\vac ) &=&
\frac{\I}{4\pi} \, \sur{x'(y)}{x(y+\gep)-x(y)-\I 0^+},
\ee
and
\beaa{rcccl}
(\vac, \wh{\gU}_\gep (y)\, \vac ) &=& 
(\vac, \gU_\gep (x(y)) \, \vac ) &=& 0,
\label{ACOU}
\eeaa
from which eq.~(\ref{COU}) follows.
\sub{Second proof of eq.~(\ref{COU})} 
\label{ap:COUnormal}
The outgoing current is computed from def.~(\ref{defCOUnormal}).
Using analogous relations to (\ref{ATEIvma}) and (\ref{ATEIvmb})
for the complex scalar field, and the equality 
\be
\intii \sur{dk}{k}\, U(k,p)\, U(k,p')^* &=& 
\sur{1}{p}\ \gd\, (p-p'),
\ee 
deduced from eq.~(\ref{commutatorequality}),
we obtain again eq.~(\ref{COU}).
\sub{Proof of eq.~(\ref{Ntotbis})} \label{ap:Ntotbis}
We follow here ref.~\cite{GaW}.
Equations (\ref{number}) and (\ref{numberbis}) imply
\be
\mbox{} \hspace{-9mm}
\intoi\sur{dk}{2k}\,B(k,p)\,B(k,p')^* 
&=& -\sur{1}{8\pi^2}\intii dy\intii dy'\,e^{\I py}\,e^{-\I py'}\,
\log \left[\,\sur{x(y)-x(y')}{y-y'}\,\right]\cdot
\label{apNtotalbisA}
\ee
Integrating by parts, we deduce from 
eqs (\ref{Ntot}) and (\ref{apNtotalbisA})
\be
\bar{N}_o^{tot} &=&
\sur{1}{4\pi^2}\intii dy\intii dy'\ \sur{1}{y-y'+\I 0^+}\
\left[\,\sur{x'(y)}{x(y)-x(y')}-\sur{1}{y-y'}\,\right].
\label{apNtotalbisB}
\ee
The expression in the square brackets is well defined in the limit
$y'\rightarrow y$:
\be
\lim_{y'\rightarrow y} 
\left[\,\sur{x'(y)}{x(y)-x(y')}-\sur{1}{y-y'}\,\right] 
&=& \partial_y\log\sqrt{x'(y)}\cdot
\label{apNtotalbisC}
\ee
The double-integral (\ref{apNtotalbisB}) contains the imaginary 
contribution $\I \pi\,\gd(y-y')$ whose regularized integral vanishes,
\be
\left.\I\,\log\sqrt{x'(y)}\ e^{-\gep\vert y\vert}\,
\right\vert_{-\infty}^{+\infty} &=& 0,
\ee 
where $\gep>0$.
Equation (\ref{apNtotalbisB}) then implies the result (\ref{Ntotbis}).
\sub{Proof of eq.~(\ref{Nnparticles})}
\label{ap:Nnparticles}
By definition
\be
\bar{N}_o[f^{(n)}] &=& (\vac, \ch{\phi}[f_n]^\dagger \, ... \,
\ch{\phi}[f_2]^\dagger \, \ch{\phi}[f_1]^\dagger \,
\ch{\phi}[f_1] \, \ch{\phi}[f_2] \, ... \, \ch{\phi}[f_n]\,\vac).
\label{ANnparticlesa}
\ee
Defining
\be
\rond{f}_i &=&\left\{ 
\begin{array}{cl}
f_i, & i= 1,2,...,n, \\[2mm]
f^*_{i-n}, & i=n+1,n+2,...,2n,
\end{array} \right. 
\ee
and assuming $\bra f_i,f_j\ket= \gd_{ij}$, we deduce from theorem
\ref{th:formulae} and for the real scalar field:
\be
(\vac,\ch{\phi} [\indice{\rond{f}}{i}] \, 
\ch{\phi} [\indice{\rond{f}}{j}]\, \vac) &=&
G\, (\indice{\rond{f}}{i}\times \indice{\rond{f}}{j} ),
\saut i,j = 1,2,...,2n.
\label{ANnparticlesb}
\ee
Using Wick's theorem, we obtain from eqs (\ref{ANnparticlesa}) and
(\ref{ANnparticlesb}) the result
\be
\bar{N}_o[f^{(n)}] &=& 
\sur{C^2}{n!\, 2^n} \,\SSum_{\tau\in {\cal P}_{2n}}
G\, (\indice{\rond{f}}{\tau(1)}\times \indice{\rond{f}}{\tau(2)})\
\ ... \ 
G\, (\indice{\rond{f}}{\tau(2n-1)}\times\indice{\rond{f}}{\tau(2n)}), \ \
\ee
which is equivalent to eq.~(\ref{Nnparticles}).
\sub{Proof of eq.~(\ref{thermalfieldop})}
\label{ap:thermalfieldop}
I follow here ref.~\cite{FR}.
The physical system is restricted to the interval $[-L,L]$.
The partition function $Z_{_L}$ is given by
\beaa{rcccl}
Z_{_L} &=& \SSum_{n_1,n_2,...=0}^\infty 
\exp\left[\,-\gb\SSum_{k=1}^\infty n_k \go_k\,\right]
&=& \Prod_{k=1}^\infty \sur{1}{1-e^{-\gb \go_k}},
\label{apthermalfieldopA}
\eeaa 
and is IR divergent in the limit $L\rightarrow\infty$.
We have furthermore
\beaa{rcccc}
\hspace{-5mm}
\Tr_{_L} \left[ \ e^{-\gb H_L} \ 
a_i^\dagger \, a_j \ \right]
&=& \SSum_{n_1,n_2,...=0}^\infty 
\exp\left[\,-\gb\SSum_{k=1}^\infty n_k\,\go_k\,\right] 
\, 2n_i\,\go_i\,\gd_{i,j} && 
\label{apthermalfieldopB}  \\ [1mm]
&=& -Z_{_L}\, (1-e^{-\gb \go_i}) \,
\sur{\partial}{\partial\gb} \,
\SSum_{n_i=0}^\infty e^{-\gb n_i \go_i} \,2 \gd_{i,j}
&=& Z_{_L}\,\sur{2\go_i}{e^{\gb \go_i}-1} 
\ \gd_{i,j}.
\eeaa 
Equations (\ref{apthermalfieldopA}) and (\ref{apthermalfieldopB})
imply the result (\ref{thermalfieldop}) in the thermodynamic limit.
\sub{Proof of eq.~(\ref{thermalfieldopn})}
\label{ap:thermalfieldopn}
We assume for simplicity that $n=2$.
Similar computations to those of appendix \ref{ap:thermalfieldop} 
lead to the result
\be
Z^{-1}_{_L}\ \Tr_{_L} 
\left[ \, e^{-\gb H_{_L}} \, 
a_i^\dagger \, a_j^\dagger\, 
a_k \, a_l\, \right] &=& 
\sur{4 \go_i \,\go_j}{(e^{\gb \go_i}-1)\,(e^{\gb \go_j}-1)} \ 
( \, \gd_{i,l} \,\gd_{j,k} + \gd_{i,k}\,\gd_{j,l}  \, ),
\label{apthermalfieldopnA}
\ee
which is also valid in the particular case $i=j=k=l$. 
In the thermodynamic limit, eq.~(\ref{thermalfieldopn}) 
is then deduced for $n=2$. 
Note that a hypothetical supplementary term like $\gd_{i,j,k,l}$ 
in eq.~(\ref{apthermalfieldopnA}) could not survive
in the thermodynamic limit.
\sub{Proof of eqs (\ref{periodic2pfthermal}) and (\ref{thermalsomme})}
\label{ap:thermalsomme}
Equation (\ref{periodic2pfthermal}) is deduced from 
def.~(\ref{moyennethermique}) using the cyclic property of the trace.
To prove eq.~(\ref{thermalsomme})
the physical system is restricted to the interval $[-L,L]$, for which
the thermal two-point function (\ref{defthermal2pf}) will be denoted by 
$W_{\gb,L}^{Th}(t,t')$.
Equations (\ref{fieldL}) and (\ref{Thermalcommutators}) show that
\be
W_{\gb,L}^{Th}(\tau,\tau')
&=& \sur{1}{2L}\sum_{i=1}^\infty \sur{1}{2\go_i}\,
\left[\  \sur{e^{\I\go_i(\tau-\tau')}}{e^{\gb\go_i}-1}
+\sur{e^{\I\go_i(\tau'-\tau)}}{1-e^{-\gb\go_i}} \ \right],
\label{apthermalsommeA}
\ee
where we have used the discretized version of eqs (\ref{thermalfieldop})
to (\ref{thermalfieldopter}).
Noting that
\be 
\sur{1}{1-e^{-\gb\go_i}} &=&  
\SSum_{n=0}^\infty e^{-n\gb\go_i},
\label{apthermalsommeB}
\ee 
we obtain
\be
W_{\gb,L}^{Th}(\tau,\tau')
&=& \sur{1}{2L} \sum_{i=1}^\infty \sur{1}{2\go_i\,}\,
\left[\, \sum_{n=1}^\infty e^{\I \go_i\,(\tau-\tau'+\I n\gb)} +
\sum_{n=0}^\infty  e^{-\I \go_i\,(\tau-\tau'-\I n\gb)} \,\right],
\label{apthermalsommeC}
\ee
from which we deduce
\be
\Re W_{\gb,L}^{Th}\, (\tau,\tau') 
&=& \sum_{n=-\infty}^{+\infty}
\Re W_{\infty,L}^{Th}\,(\tau+\I n\gb,\tau').
\label{apthermalsommeD}
\ee
We obtain the result (\ref{thermalsomme}) by taking the thermodynamic 
limit of this last equation.
\sub{Proof of eq.~(\ref{2pfthermal})} \label{ap:2pfthermal}
We define the primitives of $f_i\in\so$ as
$F_i(t)=\int_{-\infty}^t dt'\, f_i(t')$, $i=1,2$.
Integrating eq.~(\ref{defthermal2pfter}) twice by parts we obtain
\be
\mbox{}\hspace{-10mm}
W^{Th}_\gb[\,f_1\times f_2^*\,] 
&=& \sur{1}{4\pi}\intii d\tau \intii d\tau'\, F_1(\tau)\, F_2(\tau')^* 
\intii d\go \, e^{\I \go\,(\tau'-\tau)}\,\sur{\go}{1-e^{-\gb\go}}\cdot
\label{ap:2pfthermal1}
\ee
We interpret $\tau'$ as $\tau'+\I 0^+$ to regularize this integral.
Using the formulae \cite{Ob}
\be
\mbox{}\hspace{-11mm}
2\intoi d\go \, \cos\, [\,\go\,(\tau'-\tau)\, ] \ 
\sur{\go}{e^{\gb\go}-1} 
&=& \sur{1}{(\tau'-\tau)^2}- 
\left( \sur{\pi}{\gb}\right)^2 \ 
\sur{1}{\sinh^2 \,[\,\gb\,(\tau'-\tau)/\pi\, ]}, 
\\ [2mm] \mbox{}\hspace{-11mm}
\intii d\go \, \sin\, [\, \go\,(\tau'-\tau)\, ] \,
\frac{\go}{1-e^{-\gb\go}}
&=& \intoi d\go \, \sin\, [\, \go\,(\tau'-\tau)\, ] \, \go,
\ee
we deduce from eq.~(\ref{ap:2pfthermal1})
\be
W^{Th}_\gb[f_1\times f_2^*]
&=& \sur{1}{4\pi}\intii d\tau \intii d\tau'\ 
F_1(\tau)\ F_2(\tau')^* \intoi d\go\ e^{\I \go\,(\tau'-\tau)}\ \go
\hspace{30mm}
\ee
\be 
\nonumber \\ [-6mm] \nonumber
+\ \sur{1}{4\pi}\intii d\tau \intii d\tau'\
F_1(\tau)\ F_2(\tau')^* \ \partial_\tau \, \partial_{\tau'}
\left\{\,\log (\tau'-\tau)
-\log \sinh\left[\, \sur{\pi}{\gb} \,(\tau'-\tau)\, \right]\,
\right\}.
\ee
Performing again a double integration by parts we obtain
\be
\mbox{}\hspace{-6mm}
W^{Th}_\gb[f_1\times f_2^*] &=&
- \sur{1}{4\pi}
\intii d\tau\intii d\tau'\ f_1(\tau)\ f_2(\tau')^* \
\log \sinh \left[\, \sur{\pi}{\gb} \,(\tau'-\tau)\,\right].
\ee
The kernel $W_\gb^{Th}(\tau,\tau')$ is contained in this double 
integral.
The arbitrary constant is chosen so as to obtain the 
expression (\ref{Wightmanxunibis}) for the two-point function in the
limit $\gb \rightarrow \infty$.
\sub{Proof of eq.~(\ref{UkernelTNR})} 
\label{ap:UnoyauTNR}
Definition (\ref{kernelV}) and eq.~(\ref{transfoTNR}) imply that
the kernel of $V$ is given by
\be
V(k,p) &=&
e^{-\I k\gD}\ V_o\left(k,\sur{p}{M}\right)
\label{ap:V1}
\ee
where we have defined
\be
V_o(k,p) &=& \frac{1}{2\pi} 
\int_0^\infty dx \, e^{\I kx} \,x^{-\I p}.
\label{ap:V2}
\ee
Changing to the variable $s=\vert k\vert\, x$ we obtain
\be
V_o(k,p) &=& \frac{1}{2\pi}\,\sur{1}{\abs{k}^{1-\I p}} \, 
\left[\ \gt(k) \, J(-p) + \gt(-k) \, J(p)^* \ \right],
\label{ap:V}
\ee
where we have defined
\be
J(p) &=& \int_0^\infty ds \, e^{\I s} \,s^{\I p}.
\label{integralJ}
\ee
This integral is computed by deforming the contour along $\R_+$ to 
the imaginary positive axis:
\be
J(p) &=& -p \ e^{-\frac{\pi}{2} p} \ \gC (\I p).
\ee 
Since
\be
\left\vert \,\gC(\I p)\,\right\vert^2 &=& \sur{\pi}{p \, \sinh (\pi p)},
\label{identitygamma}
\ee 
we obtain
\be
J(p) &=& -p\ e^{\I \gO(p)} \ \sqrt{\sur{2\pi}{p\,(e^{2\pi p}-1)}},
\label{ap:J}
\ee  
where we have defined $\gO(p) = {\rm Arg}\,[\gC(\I p)]$.
Equations (\ref{ap:V}) and (\ref{ap:J}) show that
\be
V_o(k,p) &=& \sur{p}{\sqrt{2\pi}} \,
\sur{e^{-\I \gO(p)}\,e^{\I p \log \vert k \vert}}{\abs{k}} \
\left[ \ \sur{\gt(k)}{\sqrt{p\,(1-e^{-2\pi p})}}
- \sur{\gt(-k)}{\sqrt{p\,(e^{2\pi p}-1)}} \ \right],
\label{eq:horizonapplicationV}
\ee
from which eq.~(\ref{UkernelTNR}) is deduced using 
eqs (\ref{eq:applicationVA}) and (\ref{ap:V1}).
\sub{Proof of eq.~(\ref{NTNR})} \label{ap:NTNR}
Equation (\ref{number}) is rewritten in the form
\be
\bar{N}_o[f] &=& 
\intoi dp \,\tf{f}(p)\intoi dp'\,\tf{f}(p')^* \,
\intoi\sur{dk}{2k}\,U(-k,p)\ U(-k,p')^*.
\label{apnumber}
\ee
Using the expression (\ref{UkernelTNR}) for the kernel of $U$ and the 
formula
\be
\intoi \frac{dk}{k}\ e^{\I \frac{(p-p')}{M}\,\log k }
&=& 2\pi M \, \gd (p-p'),
\ee
eq.~(\ref{NTNR}) is easily obtained from eq.~(\ref{apnumber}).
\sub{Proof of eq.~(\ref{Ngeneral})}
\label{ap:Ngeneral}
Using theorem \ref{th:formulae} and eq.~(\ref{UkernelTNR}) we obtain
\beaa{rcccl}
G \, (f_i \times f_j) &=& 
\bra A^*\ind{\tf{f}}{j}^*,B\ind{\tf{f}}{i} \ket &=& 0,
\mbox{\hspace{30mm}}
\label{apgeneralzero}  
\eeaa
\mbox{}\vspace{-13mm} \\
\beaa{rcccl}
G \, (f_i \times f_j^*) 
&=& \ \bra B\ind{\tf{f}}{j},B\ind{\tf{f}}{i} \ket \
&=& \intoi \sur{dp}{2p} 
\sur{\ind{\tf{f}}{j}(p)^*\ind{\tf{f}}{i}(p)}{e^{\frac{2\pi}{M} p}-1},
\label{apgeneral}
\eeaa
where $i,j=1,2,...,n$.
Expression (\ref{apgeneralzero}) vanishes because of
the presence in its kernel of the term $\gd(p+p')$. 
Theorem \ref{th:Nnparticles} then implies that
\be 
\bar{N}_o[f^{(n)}] 
&=& C^2\SSum_{\gs \in {\cal P}_n}\ 
G(f_1\times f_{\gs(1)}^*) \ G(f_2\times f_{\gs(2)}^*) \ ...\ 
G (f_n\times f_{\gs(n)}^*). \ \ 
\ee
Noting that
$G\,(f_i\times f_j^*) = G_{\frac{2\pi}{M},out}^{Th} (f_i\times f_j^* )$ 
(see eq.~(\ref{Gthermal})), eq.~(\ref{Ngeneral}) is deduced from 
theorem~\ref{th:thermiqueNgeneral}.
\end{document}